# Exact duality and local dynamics in SU(N) lattice gauge theory

Manu Mathur[*] and Atul Rathor[†]

S. N. Bose National Centre for Basic Sciences, Block JD, Sector III, Salt Lake, Kolkata 700 106, India

We construct exact duality transformations in pure SU(N) Hamiltonian lattice gauge theory in $(2+1)$ dimension. This duality is obtained by making a series of iterative canonical transformations on the SU(N) electric vector fields and their conjugate magnetic vector potentials on the four links around every plaquette. The resulting dual description is in terms of the magnetic scalar fields or plaquette flux loops and their conjugate electric scalar potentials. Under SU(N) gauge transformations they both transform like adjoint matter fields. The dual Hamiltonian describes the nonlocal self-interactions of these plaquette flux loops in terms of the electric scalar potentials and with inverted coupling. We show that these nonlocal loop interactions can be made local and converted into minimal couplings by introducing SU(N) auxiliary gauge fields along with new plaquette constraints. The matter fields can be included through minimal coupling. The techniques can be easily generalized to $(3+1)$ dimensions.

## I. INTRODUCTION

In the past few decades there have been numerous approaches to dualize gauge theories to obtain their dynamics in terms of the dual potentials [1–3]. Many of these attempts are partly inspired by the success of the dual formulation of Abelian lattice gauge theories where duality transformations have led to interesting confining and nonconfining phases in terms of the magnetic monopoles [3]. It is widely believed that color confinement and nonperturbative vacuum structure can also be better understood within the dual framework with inverted couplings [1,2,4,5]. In the recent past, the quest for quantum simulation of non-Abelian lattice gauge theory Hamiltonians using trapped ion or ultra cold atomic gases and optical lattices are important and exciting developments [6–8]. The present work with local dual interactions and inverted coupling provides an alternative framework for these quantum simulations in the magnetic basis [6,8]. In fact, dual Hamiltonian formulations and the corresponding magnetic basis are of importance for quantum simulations of gauge theories as they are expected to be more cost efficient for the Hilbert space truncation processes in the weak coupling continuum limit [8]. For this reason in the last few years there has been a surge in the search for the dual representation of various Abelian and non-Abelian lattice models and their application to quantum computations [6,8,9]. The exact duality transformations also naturally lead us to the construct the dual magnetic disorder operators [9,10], which in turn, have been used in $Z_N$ and $SU(N)$ toric code models to construct anyonic states for topological quantum computations [11].

All duality approaches in the past focus on solving the Abelian or non-Abelian Gauss law constraints to write the electric fields in terms of the dual electric potentials. In Abelian gauge theories such solutions are simple and lead to interesting dynamics [3,8]. However, in non-Abelian cases the duality attempts have not been very successful. Various solutions of non-Abelian Gauss laws lead to the dual descriptions of dynamics, which are involved [1,2,5] and often nonlocal [9] with difficult physical interpretations. These nonlocal interactions also make them computationally unwieldy. Further, many of these duality techniques are tailor made for the SU(2) gauge group [5] and their generalizations to SU(3) and higher SU(N) groups are far from clear. In this paper, using a Hamiltonian approach in $(2+1)$ dimension, we illustrate how to evade the above difficulties and transit from the original SU(N) Kogut-Susskind electric vector field and magnetic vector potential description [see (1)] to the (dual) magnetic scalar field and electric scalar potential description [see (49)]. The dual formulation is also a loop formulation as the dual operators involved are untraced Wilson loops over plaquettes or equivalently the magnetic fields (see Fig. 1) and their conjugate electric scalar potentials. Under SU(N) gauge transformations they both

---

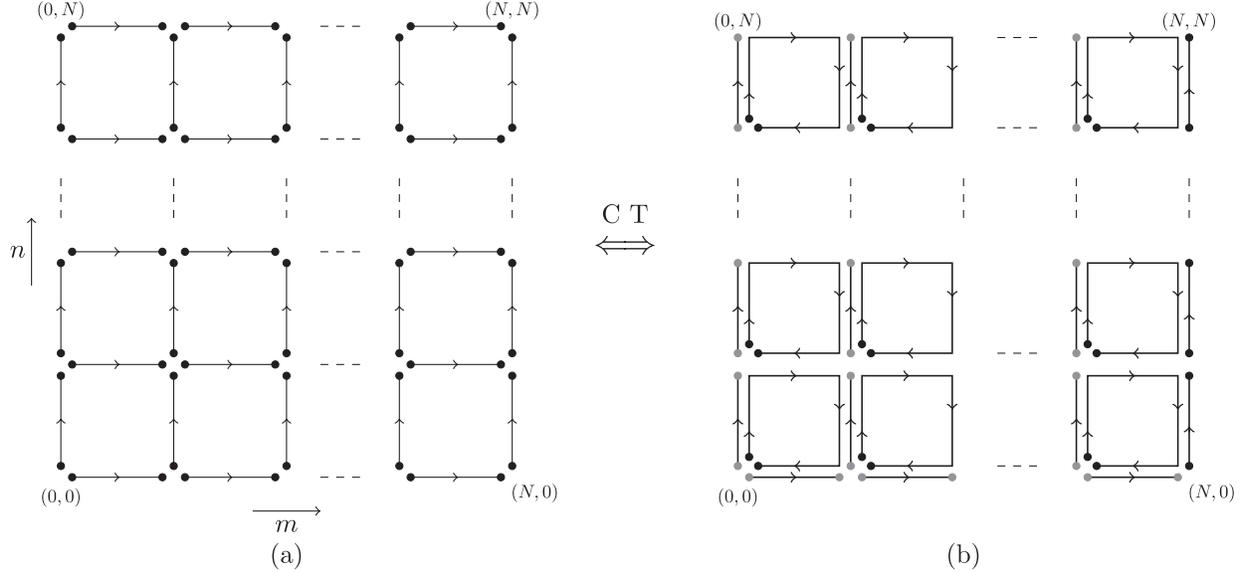

FIG. 1. (a) Original lattice with $2N(N+1)$ link holonomies and their conjugate electric fields. (b) The final configurations: $N^2$ plaquette loop holonomies, $N(N+1)$ vertical link holonomies and $N$ horizontal links holonomies at $(m, n = 0)$. The missing $N^2$ horizontal links at $(m, n > 0)$ have been traded off for $N^2$ plaquettes through canonical transformations $(CT)$ in (28) and (29). As expected, the total number of new configurations ($= N^2 + N(N+1) + N = 2N(N+1)$) in (b) match with the total number of initial link configurations in (a).

transform like adjoint matter fields. We find that the nonlocal loop-loop interactions, described by electric scalar potentials, can be made local and converted into minimal coupling by introducing auxiliary SU(N) gauge fields through additional plaquette constraints [see (41)]. This should be contrasted with the original interactions which are in terms of the magnetic vector potential holonomies around the plaquettes [see (1)]. This duality between the original plaquette link interactions and the minimal coupling interactions describing loops in $(2+1)$ dimension is a novel feature of the present study. In our previous work [9] we have constructed duality transformations that explicitly solved the SU(N) Gauss law constraints at every lattice site. The dual theory in this case was a SU(N) spin model without any gauge degrees of freedom. The above solutions of SU(N) Gauss law constraints are essentially nonlocal relations between the SU(N) Kogut-Susskind electric fields and the dual electric scalar potentials leading to a nonlocal dual Hamiltonian [9]. These nonlocality issues in the dual formulations have been recently discussed in the context of quantum simulations in the magnetic basis (see Bauer *et al.* in [6,8]). In the present work we take a different route and define the dual SU(N) electric scalar potentials without solving the Gauss law constraints. We construct SU(N) magnetic scalar or plaquette fields and their conjugate electric scalar potentials [2] by making a series of iterative canonical transformations on the original electric vector fields and their conjugate magnetic vector potentials [12]. These canonical transformations are designed to produce local plaquette loop holonomies (physical magnetic fields) by gluing together its four link holonomies (gluons). This framework is pictorially illustrated in Figs. 1 and 4. Following this process we find the following:

(1) The Kogut-Susskind noninteracting electric field $g^2 \vec{E}^2$ terms dualize to loop interaction terms. As expected, these loop interactions are described by minimal coupling between SU(N) electric scalar potentials and the corresponding auxiliary gauge fields.
(2) The Kogut-Susskind interacting magnetic field $1/g^2 \text{Tr} \ (U_{\text{plaquette}} + \text{H.c.})$ terms dualize to the noninteracting magnetic fields terms. They create and annihilate single plaquette loops [9,13].

Thus under duality the roles of interacting and noninteracting terms get interchanged resulting in the inversion of coupling constant $(g^2 \to 1/g^2)$ as expected.

The plan of the paper is as follows: In Sec. II we start with Kogut-Susskind Hamiltonian formulation. This section is added for the sake of completeness and to set up the notations. In Sec. III we discuss the canonical transformations which take us from link description to the plaquette loop description by joining the four links of every plaquette. To make the presentation simple, we first discuss how to join two link holonomies by making a single canonical transformation. In Sec. III A we iterate this step on a simple $2 \times 2$ plaquette lattice and define four new plaquette loop holonomies (magnetic fields) and their conjugate electric scalar potentials. In Sec. III B we directly generalize these results to $N \times N$ plaquette lattice and define $N^2$ new plaquette holonomies. All technical issues and details involved in performing canonical transformations are worked out in Appendix A. In Sec. IV we discuss the dual loop dynamics in (2 + 1) dimensions in terms of magnetic scalar fields and their conjugate electric scalar potentials. The nonlocality and rotational symmetry problems and their resolutions are discussed. We also compare our SU(N) duality results with the U(1) lattice gauge theory duality results. This simple comparison provides better understanding of the non-Abelian duality relations between electric fields and the electric scalar potentials. We end the paper with a summary and a brief discussion about the future problems.

The notations used are as follows: The lattice sites and links will be denoted by $\vec{n} = (m,n)$ and $(\vec{n}; \hat{i})$, respectively, with $m, n = 0, 1, 2, ..., N$ and $i = 1, 2$. We use roman and calligraphic fonts to denote the SU(N) conjugate field operators in the electric (before duality) and the magnetic (after duality) descriptions, respectively. This is clearly illustrated in Table I.

## II. THE HAMILTONIAN DYNAMICS

The Hamiltonian of SU(N) lattice gauge theory is [14]

$$H = H_E + H_M = g^2 \sum_{\vec{n},\hat{i}} \text{Tr}\,(E^2(\vec{n};\hat{i}))$$
$$+ \frac{K}{g^2} \sum_{\text{p}} \text{Tr}\,(2 - (U_{\text{p}} + U_{\text{p}}^\dagger)), \quad (1)$$

where $g^2$ is the coupling and $K$ is a constant. The plaquette operator $U_{\text{p}}$ at site $\vec{n}$ and $(i, j)$ plane is defined as $U_{\text{p}} = U(\vec{n};\hat{i})U(\vec{n}+\hat{i};\hat{j})U^\dagger(\vec{n}+\hat{j};\hat{i})U^\dagger(\vec{n};\hat{j})$ and $E_\pm = E_\pm^a T^a$, where $T^a, a = 1, 2, ..., N^2 - 1$ are the generators of fundamental representation of SU(N). This is the electric field description with the conjugate pairs $(E^a(\vec{n};\hat{i}), U(\vec{n};\hat{i}))$ satisfying

$$[E_+^a(\vec{n};\hat{i}), U_{\alpha\beta}(\vec{n};\hat{i})] = -(T^a U(\vec{n};\hat{i}))_{\alpha\beta},$$
$$[E_-^a(\vec{n}+\hat{i};\hat{i}), U_{\alpha\beta}(\vec{n};\hat{i})] = (U(\vec{n};\hat{i})T^a)_{\alpha\beta}. \quad (2)$$

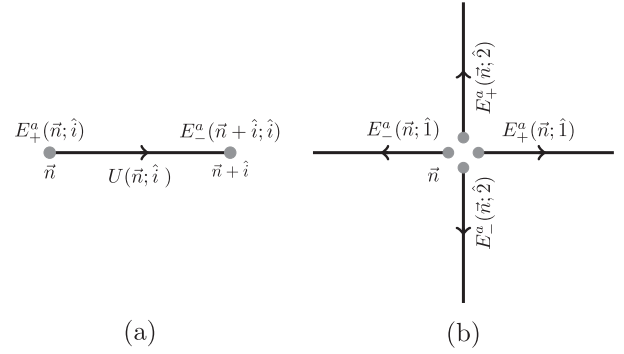

FIG. 2. Kogut-Susskind link operators $U(\vec{n};\hat{i})$ and their left (right) electric field $E_+^a(\vec{n};\hat{i})(E_-^a(\vec{n}+\hat{i};\hat{i}))$; (a) location of electric fields on the link $(\vec{n};\hat{i})$, (b) location of electric fields around a lattice site $\vec{n}$. The SU(N) Gauss law (8) involves all four electric fields around site $\vec{n}$.

In (1) we have used $\text{Tr}\,E^2(\vec{n};\hat{i}) \equiv \text{Tr}\,E_\pm^2(\vec{n};\hat{i})$. The above commutation relation implies that $E_+^a(\vec{n};\hat{i})$ $(E_-^a(\vec{n}+\hat{i};\hat{i}))$ shown in Fig. 2(a) rotate (antirotate) $U(\vec{n};\hat{i})$ the form left (right), respectively. The link operators $U(\vec{n};\hat{i})$ are SU(N) operators and satisfy

$$U^\dagger U = \mathcal{I} = UU^\dagger,$$

where $\mathcal{I}$ is $N \times N$ identity operator. Moreover matrix elements of $U$ commute among themselves

$$[U_{\alpha\beta}, U_{\gamma\delta}] = 0, \qquad [U_{\alpha\beta}, U_{\gamma\delta}^\dagger] = 0. \quad (3)$$

The left and right electric fields $E_\pm^a(\vec{n};\hat{i})$ commute with each other and individually satisfy SU(N) Lie algebra.

$$[E_\pm^a(\vec{n};\hat{i}), E_\pm^b(\vec{n};\hat{i})] = if^{abc} E_\pm^c(\vec{n};\hat{i}). \quad (4)$$

The right electric field $E_-^a(\vec{n}+\hat{i};\hat{i})$ rotating the link operator from the right in (2) are obtained by the parallel transports along the link $(\vec{n},\hat{i})$:

TABLE I. The kinematical degrees of freedom before and after duality transformations. Under SU(N) gauge transformations, the loop conjugate pairs $(\mathcal{E}_+^a(\vec{n}), \mathcal{W}_{\alpha\beta}(\vec{n}))$ in [B] transform like SU(N) adjoint scalar matter fields. They describe the SU(N) magnetic fields and their conjugate electric scalar potentials, respectively. The last column shows the auxiliary SU(N) gauge fields defined on links. They are introduced with additional plaquette constraints (41) to obtain minimally coupled local dual theory. This table also explains the notations used in this paper. The locations of different holonomies and their electric fields are shown in Fig. 6.

| SU(N) Kogut-Susskind formulation | Dual SU(N) formulation | |
|---|---|---|
| [A] (Mixed) | [B] (Physical) | [C] (Unphysical) |
| Link holonomy: $U_{\alpha\beta}(\vec{n};\hat{i})$ | Plaquette holonomy: $\mathcal{W}_{\alpha\beta}(\vec{n})$ | String holonomy: $\mathcal{U}(\vec{n};\hat{i})$ |
| Link electric field: $E_\pm^a(\vec{n};\hat{i})$ | Plaquette potential: $\mathcal{E}_\pm^a(\vec{n})$ | String electric field: $\mathcal{E}_\pm^a(\vec{n};\hat{i})$ |
| $[E_\pm^a(\vec{n};\hat{i}), U_{\alpha\beta}(\vec{n};\hat{i})] = -(T^a U(\vec{n};\hat{i}))_{\alpha\beta}$ | $[\mathcal{E}_\pm^a(\vec{n}), \mathcal{W}_{\alpha\beta}(\vec{n})] = -(T^a \mathcal{W}(\vec{n}))_{\alpha\beta}$ | $[\mathcal{E}_\pm^a(\vec{n};\hat{i}), \mathcal{U}_{\alpha\beta}(\vec{n};\hat{i})] = -(T^a \mathcal{U}(\vec{n};\hat{i}))_{\alpha\beta}$ |

$$E_-(\vec{n}+\hat{i};\hat{i}) = -U^\dagger(\vec{n};\hat{i})E_+(\vec{n};\hat{i})U(\vec{n};\hat{i}). \quad (5)$$

Note that Tr $(E_+(\vec{n};\hat{i}))^2 = $ Tr $(E_-(\vec{n}+\hat{i};\hat{i}))^2$ and

$$[E_+^a(\vec{n};\hat{i}), E_-^b(\vec{n}+\hat{i};\hat{i})] = 0. \quad (6)$$

Under gauge transformation at site $\vec{n}$, the left electric field and the holonomy transform as

$$U(\vec{n};\hat{i}) \to \Lambda(\vec{n})U(\vec{n};\hat{i})\Lambda^\dagger(\vec{n}+\hat{i}),$$
$$E_+(\vec{n};\hat{i}) \to \Lambda(\vec{n})E_+(\vec{n};\hat{i})\Lambda^\dagger(\vec{n}). \quad (7)$$

In (7), $\Lambda(\vec{n})$ are arbitrary unitary matrices. The SU(N) Gauss laws at the lattice site $\vec{n}$ are [14]

$$\mathcal{G}^a(\vec{n}) \equiv \sum_{i=1}^{2}(E_+^a(\vec{n};\hat{i}) + E_-^a(\vec{n};\hat{i})) = 0, \quad \forall\ \vec{n}. \quad (8)$$

## III. CANONICAL TRANSFORMATIONS: LINKS TO LOOPS AND STRINGS

In this section, using canonical transformations, we transit from the Kogut-Susskind link electric field representation to its dual plaquette magnetic field representation in SU(N) lattice gauge theory. These transformations are used to write the Hamiltonian in (1) in its dual form (49). This duality is achieved by canonical gluing the four links around every plaquette on the lattice to define plaquette loop or magnetic operators and their conjugate electric scalar potentials. This is pictorially shown in Figs. 4(a) and 4(b). Note that no attempt is made to solve the SU(N) Gauss laws explicitly to obtain this dual magnetic description. As the above canonical transformation procedure is iterative, we start with gluing two link holonomies and define their electric fields. We then generalize this canonical transformation procedure to $2 \times 2$ plaquette lattice (see Sec. III A) and then to $N \times N$ plaquette lattice (see Sec. III B), respectively. In what follows, we will construct only left (right) plaquette and string electric fields through canonical transformations. Their right (left) electric fields can then be easily obtained using the parallel transport relations (5) with $U(\vec{n};\hat{i})$ replaced by the corresponding plaquette or string holonomies. We use calligraphic symbols to denote the new field operators obtained after every canonical transformation.

We consider any two adjacent conjugate pairs: $(E_\pm^a(1), U_{\alpha\beta}(1))$ and $(E_\pm^a(2), U_{\alpha\beta}(2))$. They are the two conjugate pairs located on the links $(\vec{n};\hat{1})$ and $(\vec{n}+\hat{1};\hat{1})$, respectively, as shown in Fig. 3. More precisely,

$$(E_\pm^a(1), U_{\alpha\beta}(1)) \equiv (E_\pm^a(\vec{n};\hat{1}), U_{\alpha\beta}(\vec{n};\hat{1})),$$
$$(E_\pm^a(2), U_{\alpha\beta}(2)) \equiv (E_\pm^a(\vec{n}+\hat{1};\hat{1}), U_{\alpha\beta}(\vec{n}+\hat{1};\hat{1})).$$

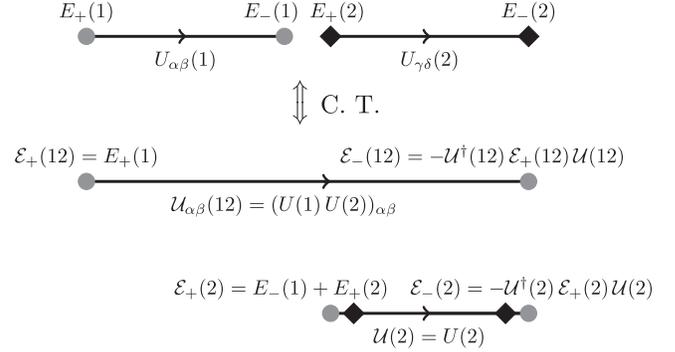

FIG. 3. Gluing two SU(N) holonomies using canonical transformations. From two Kogut-Susskind links, we get two new mutually independent holonomies.

We join the above holonomies together and define the new but equivalent pairs $(\mathcal{E}_\pm^a(12), \mathcal{U}_{\alpha\beta}(12))$ and $(\mathcal{E}_\pm^a(2), \mathcal{U}_{\alpha\beta}(2))$ through canonical transformations:

$$\mathcal{U}_{\alpha\beta}(12) \equiv (U(1)\,U(2))_{\alpha\beta}, \qquad \mathcal{E}_+^a(12) = E_+^a(1),$$
$$\mathcal{U}_{\alpha\beta}(2) \equiv U_{\alpha\beta}(2), \qquad \mathcal{E}_+^a(2) = E_-^a(1) + E_+^a(2). \quad (9)$$

The transformations (9) are illustrated in Fig. 3. They are canonical as the two new conjugate pairs $(\mathcal{E}_\pm^a(12), \mathcal{U}_{\alpha\beta}(12))$ and $(\mathcal{E}_\pm^a(2), \mathcal{U}_{\alpha\beta}(2))$ also follow the standard canonical commutation relations:

$$[\mathcal{E}_+^a(12), \mathcal{U}_{\alpha\beta}(12)] = -(T^a\mathcal{U}(12))_{\alpha\beta}, \quad (10a)$$

$$[\mathcal{E}_+^a(2), \mathcal{U}_{\alpha\beta}(2)] = -(T^a\mathcal{U}(2))_{\alpha\beta}. \quad (10b)$$

Note that the two new holonomies $\mathcal{U}_{\alpha\beta}(12)$ and $\mathcal{U}_{\alpha\beta}(2)$ trivially commute with each other and we have added $E_-^a(1)$ to define $\mathcal{E}_+^a(2)$ in (9) so that

$$[\mathcal{E}_+^a(2), \mathcal{U}_{\alpha\beta}(12)] = 0,$$
$$[\mathcal{E}_+^a(12), \mathcal{U}_{\alpha\beta}(2)] \equiv 0. \quad (11)$$

The two new conjugate pairs commute with each other and are therefore mutually independent. They are on the same footing as the original two Kogut-Susskind pairs.

Note that, in this simplest two link case, if we identify the two end points $(\vec{n})$ and $(\vec{n}+\hat{1})$ in Fig. 3 then $\mathcal{U}(12)$ transforms like a magnetic flux loop. We can now follow the classification shown in Table I by identifying $(\mathcal{E}_\pm(12), \mathcal{U}(12))$ with $(\mathcal{E}_\pm(\vec{n}), \mathcal{W}(\vec{n}))$ and $(\mathcal{E}_+(2), \mathcal{U}(2))$ with the string pair $(\mathcal{E}_+(\vec{n}+\hat{1};\hat{1}), \mathcal{U}(\vec{n}+\hat{1};\hat{1}))$. We further note that the electric field $E_\pm(1)$ of the link holonomy $U_{\alpha\beta}(1)$, which is canonically transformed into $\mathcal{U}_{\alpha\beta}(12)$, appears in both the final electric fields. This aspect is clearly shown in Fig. 3. This simple fact will lead to the nonlocal duality relations [see (14) and (20a), (20b)], which are obtained after iterating (9) over the entire lattice. This, in turn, will lead to nonlocal dual or loop dynamics [see (39)]. As mentioned earlier, having defined the left electric fields in (9), the right electric fields get fixed by the parallel transport along the new links

$$\mathcal{E}_+^a(12) = -R_b^a(\mathcal{U}(12))\,\mathcal{E}_-^b(12),$$
$$\mathcal{E}_+^a(2) = -R_b^a(\mathcal{U}(2))\,\mathcal{E}_-^b(2). \quad (12)$$

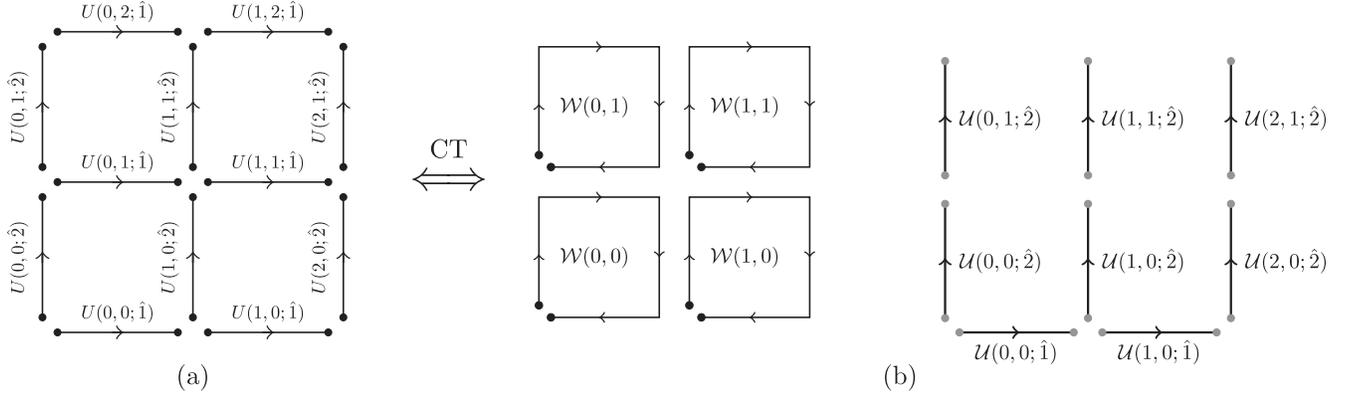

FIG. 4. A simple $2 \times 2$ plaquette lattice. (a) The Kogut-Susskind description in terms of 12 link holonomies and their left and right electric fields shown by •. (b) Dual lattice with four physical plaquette holonomies $\mathcal{W}(m,n); m, n = 0, 1$, six unphysical vertical strings $\mathcal{U}(m,n;\hat{2}); m = 0, 1, 2, n = 0, 1$, and two unphysical horizontal strings $\mathcal{U}(m,0;\hat{1}); m = 0, 1$ at the bottom of the lattice. The electric fields of plaquettes are shown by •, whereas electric fields of unphysical strings are shown in •. All unphysical strings can be removed by gauge transformations at $(m \neq 0, n \neq 0)$.

In the above equations $R_b^a$ are the $SO(N^2-1)$ rotation operators and satisfy $RR^T = 1 = R^T R$. They are defined as $R_b^a(U) = 2\mathrm{tr}(T^a U T^b U^\dagger)$. The new left and right electric fields also satisfy the SU(N) Lie algebra, they commute with each other and their magnitudes are equal. In summary, in this section we have converted the shorter flux line $U_{\alpha\beta}(1)$ into longer flux line $\mathcal{U}_{\alpha\beta}(12)$ using (9). This simple canonical transformation will now be iterated over the entire lattice to convert all horizontal links into local plaquettes starting from the top. This in turn will define the holonomy around a plaquette or the magnetic fields as the fundamental variables in the dual theory [15]. We first generalize the canonical transformations (9) to $2 \times 2$ plaquette lattice in Sec. III A and then discuss the general $N \times N$ plaquette case in Sec. III B.

### A. $(2 \times 2)$ plaquette lattice

This simple case is illustrated in Fig. 4 and in Table I. The initial 12 Kogut-Susskind link conjugate pairs $(E(m,n,\hat{i}), U(m,n;\hat{i}))$ are shown in Fig. 4(a) or in Table I[A]. The final four (physical) plaquette conjugate pairs $(\mathcal{E}(m,n), \mathcal{W}(m,n))$ and the remaining eight (unphysical) string conjugate pairs $(\mathcal{E}(m,n;\hat{i}), \mathcal{U}(m,n;\hat{i}))$ are shown in Fig. 4(b) or in Tables I[B] and I[C], respectively. As is clear from the figure, we have converted the four Kogut-Susskind horizontal link holonomies and their electric fields at $(m = 0, 1, n = 1, 2)$ into the four plaquette holonomies and their electric fields. The 12 canonical transformations leading to the configurations in Fig. 4(b) from Fig. 4(a) are systematically worked out in Appendix A. In the next section the end results of the above canonical transformations are written down. They have exact duality interpretation.

#### 1. Plaquette, strings, and duality

We first describe the new plaquette sector. The four plaquette fluxes shown in Fig. 4(b) are

$$\mathcal{W}(m,n) = U(m,n;\hat{2})\,U(m,n+1;\hat{1})$$
$$\qquad U^\dagger(m+1,n;\hat{2})\,U^\dagger(m,n;\hat{1}). \quad (13)$$

In (13) $m, n = 0, 1$. Their conjugate plaquette electric fields, fixed through the iterative canonical transformations (see Appendix A), are

$$\mathcal{E}_+(m,n) = -\sum_{j=n+1}^{N=2} \mathcal{S}_j(m,n) E_+(m+1,j;\hat{1})\mathcal{S}_j^{-1}(m,n). \quad (14)$$

The parallel transports $\mathcal{S}_j(m,n)$ in (14) are defined as (see Appendix A)

$$\mathcal{S}_{j=1}(m,0) = U(m,0;\hat{2})U(m,1;\hat{1}), \quad (15a)$$

$$\mathcal{S}_{j=2}(m,0) = U(m,0;\hat{2})U(m,1;\hat{1})U(m+1,1;\hat{2}), \quad (15b)$$

$$\mathcal{S}_{j=2}(m,1) = U(m,1;\hat{2})U(m,2;\hat{1}). \quad (15c)$$

The physical interpretation of the nonlocal operators $\mathcal{S}_j(m,n)$ shown in Figs. 5(a)–5(c) is simple. They implement the parallel transports from the location of the original Kogut-Susskind link electric fields at $(m+1, j > n)$ to the location of the dual plaquette electric field at $(m, n)$ in (14). Therefore like Kogut-Susskind electric fields the dual plaquette electric fields $\mathcal{E}(m, n)$ also transform like adjoint matter fields.

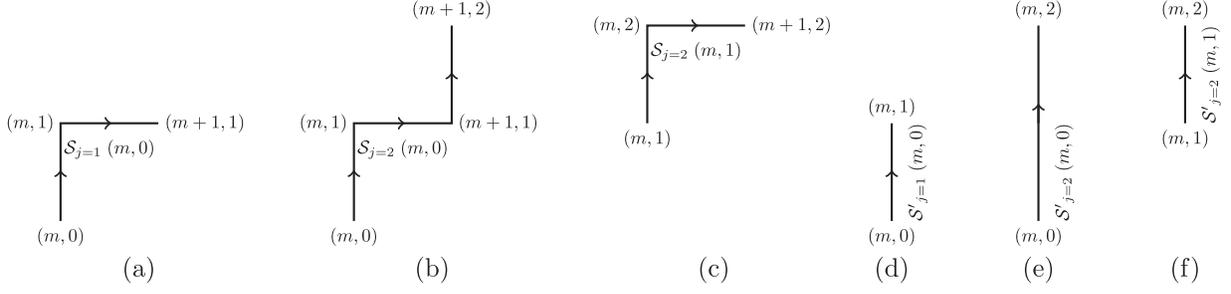

FIG. 5. Nonlocal parallel transports $\mathcal{S}$ and $\mathcal{S}'$ required for defining plaquette and string electric field operators in (15a), (15b), (15c) and (21a), (21b), (21c), respectively.

As a consequence of canonical transformations the plaquette pairs $(\mathcal{E}_{\pm}(m, n), \mathcal{W}(m, n))$ are conjugate and satisfy the standard canonical commutation relations:

$$[\mathcal{E}_+^a(\vec{n}), \mathcal{W}_{\alpha\beta}(\vec{n})] = -(T^a \mathcal{W}(\vec{n}))_{\alpha\beta},$$
$$[\mathcal{E}_-^a(\vec{n}), \mathcal{W}_{\alpha\beta}(\vec{n})] = (\mathcal{W}(\vec{n})T^a)_{\alpha\beta}. \quad (16)$$

The plaquette electric fields satisfy SU(N) Lie algebra

$$[\mathcal{E}_\pm^a(\vec{n}), \mathcal{E}_\pm^b(\vec{n})] = if^{abc}\mathcal{E}_\pm^c(\vec{n}). \quad (17)$$

The canonical commutation relation amongst the new plaquette fields (16), (17) are the dual version of the standard Kogut-Susskind commutation relations (2), (4), respectively. Note that while Kogut-Susskind relations (16) involve electric fields $E(\vec{n}; \hat{i}))$ and its conjugate magnetic vector potentials in $U_{\alpha\beta}(\vec{n}; \hat{i})$, the dual commutation relations (16) involve magnetic scalar fields in $\mathcal{W}(m, n)$ and their conjugate electric scalar potentials $\mathcal{E}(m, n)$. Therefore, the canonical transformations (13) and (14) can also be interpreted as the exact SU(N) duality transformations.

After duality, the fundamental conjugate pairs describing the dynamics (see Sec. IV) are the magnetic scalar fields and their conjugate electric scalar potentials $(\mathcal{E}(m, n), \mathcal{W}(m, n))$. Under SU(N) gauge transformations (7) they transform as adjoint scalar matter fields

$$\mathcal{W}(\vec{n}) \to \Lambda(\vec{n})\mathcal{W}(\vec{n})\Lambda^\dagger(\vec{n}),$$
$$\mathcal{E}_\pm(\vec{n}) \to \Lambda(\vec{n})\mathcal{E}_\pm(\vec{n})\Lambda^\dagger(\vec{n}). \quad (18)$$

We now describe the remaining eight unphysical string sector shown in Fig. 4(b) and Table I[C]. As the iterative canonical transformations preserve the total number of degrees of freedom, these eight strings are the leftover degrees of freedom after defining the four dual plaquette holonomies in (13). They are unphysical and can be completely gauged away as is clear from Fig. 4(b). However we retain them to keep the dual loop dynamics simple and local (see Sec. IV). The two horizontal and six vertical string holonomies are

$$\mathcal{U}(m, 0; \hat{1}) = U(m, 0; \hat{1}); \qquad m = 0, 1. \quad (19a)$$

$$\mathcal{U}(m, n; \hat{2}) = U(m, n; \hat{2}); \quad m = 0, 1, 2, \ n = 0, 1. \quad (19b)$$

The corresponding conjugate electric fields are (see Appendix A)

$$\mathcal{E}_+(m, 0; \hat{1}) = E_+(m, 0; \hat{1}) - \sum_{j=1}^{2} \mathcal{S}_j(m, 0) E_-(m+1, j; \hat{1}) \mathcal{S}_j^{-1}(m, 0),$$

$$\mathcal{E}_+(m, n; \hat{2}) = E_+(m, n; \hat{2}) - \sum_{j=n+1}^{2} \mathcal{S}'_j(m, n) E_-(m, j; \hat{1}) \mathcal{S}'^{-1}_j(m, n) \quad (20a)$$

$$+ \sum_{j=n+1}^{2} \mathcal{S}_j(m, n) E_-(m+1, j; \hat{1}) \mathcal{S}_j^{-1}(m, n). \quad (20b)$$

Again like the parallel transport $\mathcal{S}_j(m,n); (m,n=0,1)$, the parallel transports $\mathcal{S}'_j(m,n)$ in (20b) are required to construct the new string electric fields (see Fig. 5) in the dual description. They are defined as

$$\mathcal{S}'_1(m,0) = U(m,0;\hat{2}), \tag{21a}$$

$$\mathcal{S}'_2(m,0) = U(m,0;\hat{2})U(m,1;\hat{2}), \tag{21b}$$

$$\mathcal{S}'_2(m,1) = U(m,1;\hat{2}). \tag{21c}$$

The parallel transports $\mathcal{S}'_j(m,n)$ in (21a), (21b), and (21c) are shown in Figs. 5(d)–5(f), respectively. For more details, see Eqs. (A24), (A44), (A50), (A14), and (A39) in Appendix A. Again the canonical transformations ensure that the eight-string conjugate pairs also satisfy the standard canonical commutations relations:

$$[\mathcal{E}^a_+(m,n;\hat{i}), \mathcal{U}_{\alpha\beta}(m,n;\hat{i})] = -(T^a \mathcal{U}(m,n;\hat{i}))_{\alpha\beta}. \tag{22}$$

In (22) if $\hat{i} = \hat{1}$ then $n = 0$ as horizontal strings exist only at the bottom. These canonical relations are systematically derived in Appendix A. It is easy to check that the 12 new conjugate pairs commute with each other and therefore are completely independent. Thus the canonical transformations again ensure that there is no mismatch between the 12 initial (links) and the 12 final (loops and strings) degrees of freedom. While the loop conjugate pairs transform as SU(N) adjoint matter, the string conjugate pairs transform as SU(N) gauge fields

$$\mathcal{U}(\vec{n};\hat{i}) \to \Lambda(\vec{n})\mathcal{U}(\vec{n};\hat{i})\Lambda^\dagger(\vec{n}+\hat{i}),$$
$$\mathcal{E}_\pm(\vec{n};\hat{i}) \to \Lambda(\vec{n})\mathcal{E}_\pm(\vec{n};\hat{i})\Lambda^\dagger(\vec{n}). \tag{23}$$

In (23) the conjugate pairs $(\mathcal{E}_\pm(\vec{n};\hat{i}=1), \mathcal{U}(\vec{n};\hat{i}=1))$ in (23) exist only when $\vec{n} = (m,0)$. Note that this asymmetry in the string holonomy sector is due to the special choice of canonical transformations in Appendix A, which converts all Kogut-Susskind horizontal link holonomies at $(m, n>0)$ into plaquette loops. Their absence also leads to nonlocal loop-loop interactions. This is because the nearest neighbour electric scalar potentials $\mathcal{E}(\vec{n})$ and $\mathcal{E}(\vec{n}+\hat{i})$ cannot be coupled minimally in the horizontal directions (see Sec. IV). In Sec. IV A we will reintroduce the horizontal holonomies through new plaquette constraints (41) and recover the rotational symmetry as well as locality of the original Hamiltonian (1).

### 2. Inverse relations

The canonical relations (13), (19a), and (19b) can be easily inverted to write the Kogut-Susskind fields in terms of the new plaquette, string fields:

$$U(\vec{n};\hat{1}) = \mathcal{L}(\vec{n})\mathcal{U}(m,0;\hat{1})\mathcal{R}(\vec{n}+\hat{1}), \quad n = 1,2, \tag{24a}$$

$$U(m,0;\hat{1}) = \mathcal{U}(m,0;\hat{1}), \quad m = 0,1, \tag{24b}$$

$$U(m,n;\hat{2}) = \mathcal{U}(m,n;\hat{2}). \quad m = 0,1,2. \tag{24c}$$

In (24a), the parallel transports on the left and right sides are

$$\mathcal{L}(\vec{n}) \equiv \left[\prod_{j=1}^{n} \mathcal{U}^\dagger(m,n-j;\hat{2})\mathcal{W}(m,n-j)\right], \tag{25a}$$

$$\mathcal{R}(\vec{n}+\hat{1}) \equiv \left[\prod_{k=0}^{n-1} \mathcal{U}(m+1,k;\hat{2})\right]. \tag{25b}$$

Note that the nontrivial relations (24a) and (25a), (25b), involving nonlocal parallel transports $\mathcal{L}$ and $\mathcal{R}$, are again simple consequence of the covariance under the SU(N) gauge transformations (7), (18), and (23). The corresponding Kogut-Susskind electric fields are

$$E_+(m,0;\hat{1}) = \mathcal{E}_-(m,0) + \mathcal{E}_+(m,0;\hat{1}), \tag{26a}$$

$$E_+(\vec{n};\hat{1}) = \mathcal{E}_-(\vec{n}) + \mathcal{U}^\dagger(\vec{n}-\hat{2};\hat{2})\mathcal{E}_+(\vec{n}-\hat{2})\mathcal{U}(\vec{n}-\hat{2};\hat{2}), \tag{26b}$$

$$E_+(\vec{n};\hat{2}) = \mathcal{E}_+(\vec{n}) + \mathsf{S}^{-1}(\vec{n}-\hat{1};\hat{1})\mathcal{E}_-(\vec{n}-\hat{1})\mathsf{S}(\vec{n}-\hat{1};\hat{1}) + \mathcal{E}_+(\vec{n};\hat{2}). \tag{26c}$$

In (26c) nonlocal parallel transports are

$$\mathsf{S}(m,n;\hat{1}) = \mathcal{L}\mathcal{U}(m,0;\hat{1})\mathcal{R}. \tag{27}$$

These relations are derived in Appendix A. However, they are easy to understand and can be written down just by looking at the final four loop and eight string configurations in Fig. 4(b). We note that the Kogut-Susskind electric fields $E_\pm(m,n,\hat{i})$ rotates all those new configurations in Fig. 4(b) that share the link holonomy $U(m,n;\hat{i})$. Therefore, $E_\pm(m,n,\hat{i})$ is a sum of all these loop and string electric fields parallel transported to the lattice site $(m,n)$ to maintain the SU(N) gauge covariance. As an example, if we want to write the Kogut-Susskind right electric field $E_+(m,n;\hat{1})$, then we identify the two dual holonomies $\mathcal{W}(m,n)$ and $\mathcal{W}(m,n-1)$, which share the link $U(m,n;\hat{1})$ and parallel transport their electric field to the site $(m,n)$ to get the canonical relation (26b). Similarly, the Kogut-Susskind electric fields $E_+(m,0,\hat{1}), E_+(m,n,\hat{2})$ in (26a) and (26c) get extra contribution from the respective horizontal and vertical strings. These inverse relations are graphically illustrated in Figs. 8(a)–8(c).

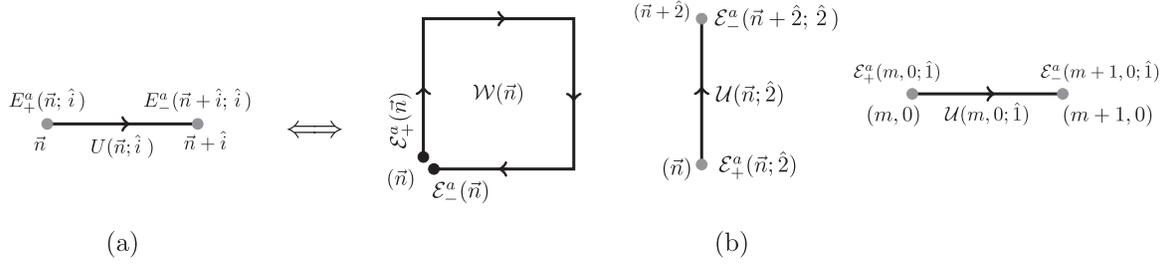

FIG. 6. (a) Kogut-Susskind Link operators, (b) $N^2$ plaquette and $N(N+1)$ vertical and $N$ horizontal string at the bottom of the lattice. The two types of electric fields $E_\pm(\vec{n};\hat{i}), \mathcal{E}_\pm(\vec{n}), \mathcal{E}_\pm(\vec{n};\hat{i})$ and their locations are shown.

### B. $(N \times N)$ plaquette lattice

We now generalize the dual relations obtained in the previous section to $N \times N$ lattice. There are $N^2$ horizontal links at $(m, n > 0)$ as shown in Fig. 1. Using (9) we canonically transform them into plaquettes in the clockwise direction as shown in Fig. 6(b). This canonical gluing starts from the top left column and goes from the top to the bottom and then repeated iteratively in the adjacent right columns. As each plaquette formation requires three canonical transformations (see Appendix A), we need $3N^2$ canonical transformations to cover the entire lattice. At the end we construct (i) $N^2$ plaquettes pairs: $(\mathcal{E}(\vec{n}), \mathcal{W}(\vec{n}))$, (ii) $N(N+1)$ vertical strings pairs: $(\mathcal{E}(\vec{n}; \hat{2}), \mathcal{W}(\vec{n}; \hat{2}))$, and (iii) $N$ horizontal string pairs: $(\mathcal{E}(m, 0; \hat{1}), \mathcal{W}(m, 0; \hat{1}))$. These dual configurations with their left, right electric fields are shown in Fig. 6(b).

#### 1. Plaquette, strings, and duality

The $N^2$ plaquettes fluxes are

$$\mathcal{W}(\vec{n}) = U(\vec{n}; \hat{2}) U(\vec{n} + \hat{2}; \hat{1}) U^\dagger(\vec{n} + \hat{1}; \hat{2}) U^\dagger(\vec{n}; \hat{2}). \quad (28)$$

Their conjugate left electric fields are

$$\mathcal{E}_+(\vec{n}) = -\sum_{j=n+1}^{N} \mathcal{S}_j(\vec{n}) E_-(m+1, j; \hat{1}) \mathcal{S}_j^{-1}(\vec{n}) \quad (29)$$

in (29) $\vec{n} \equiv (m, n)$ and $m, n = 0, 1, \ldots, N-1$. The canonical relations (29) are straightforward generalizations of the relations (14) where we have replaced 2 with $N$. This generalization amounts to including all horizontal Kogut-Susskind electric fields $E_-(m+1, j > n; \hat{1})$ up to the top of the lattice. In (29) we have defined the parallel transport operator $\mathcal{S}_j(m, n)$ shown in Fig. 7(a):

$$\mathcal{S}_j(m, n) \equiv U(m, n; \hat{2}) U(m, n+1; \hat{1}) \prod_{k=n+1}^{j-1} U(m+1, k; \hat{2}). \quad (30)$$

In (30) $j \geq n + 2$ and $\mathcal{S}_{n+1}(m, n) \equiv U(m, n; \hat{2}) U(m, n+1; \hat{1})$. The nonlocal parallel transport operators $\mathcal{S}_j(m, n)$ encode the cumulative effects of all $3N^2$ canonical transformations over the entire lattice. As mentioned in the previous section, they are necessary for SU(N) gauge covariance of (30).

The asymmetry in the shape of the $\mathcal{S}_j(m, n)$ is because of the choice of iterative canonical transformations. In this work we started at the left top corner and proceeded toward the bottom in the first column and then moved to the adjacent right column. We know that $\mathcal{W}(m, n); n = (N - 1), (N - 2), \cdots 0$ are created sequentially by absorbing $U(m, n+1; \hat{1})$ at $(N - n)$th step starting from the top. Therefore its electric field must contain all $(N - n)$ Kogut Susskind electric fields on the horizontal links above it. They are located at different points and are parallel transported to $(m, n)$ via path $\mathcal{S}$ to maintain gauge covariance of (29). The plaquette canonical commutation relations (16) and (17) discussed in the previous section on the simple $2 \times 2$ lattice remain valid.

Having discussed the plaquette loop or magnetic field sector, we now discuss the remaining string sector. The $N$ horizontal and $N(N+1)$ vertical strings are related to old link variables as

$$\mathcal{U}(m,n=0;\hat{1}) = U(m,0;\hat{1}), \quad \mathcal{U}(\vec{n};\hat{2}) = U(\vec{n};\hat{2}). \tag{31}$$

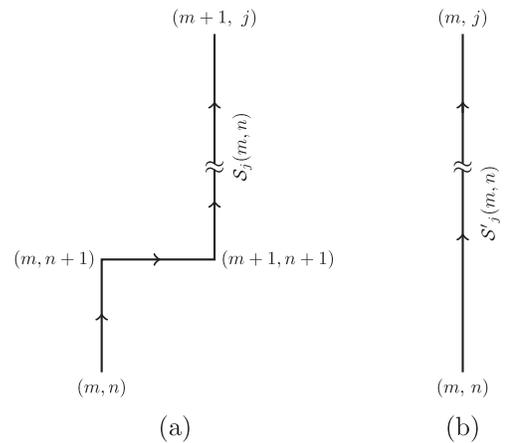

FIG. 7. Two types of nonlocal parallel transports required for canonical transformation: (a) $\mathcal{S}_j(m, n)$ defining the electric fields of $\mathcal{W}(m, n)$ in (29) and (32b), (b) $\mathcal{S}'_j(m, n)$ defining the electric fields of the strings $\mathcal{U}(m, n; \hat{i})$ in (32b). These strings are $N \times N$ lattice generalizations of Fig. 5.

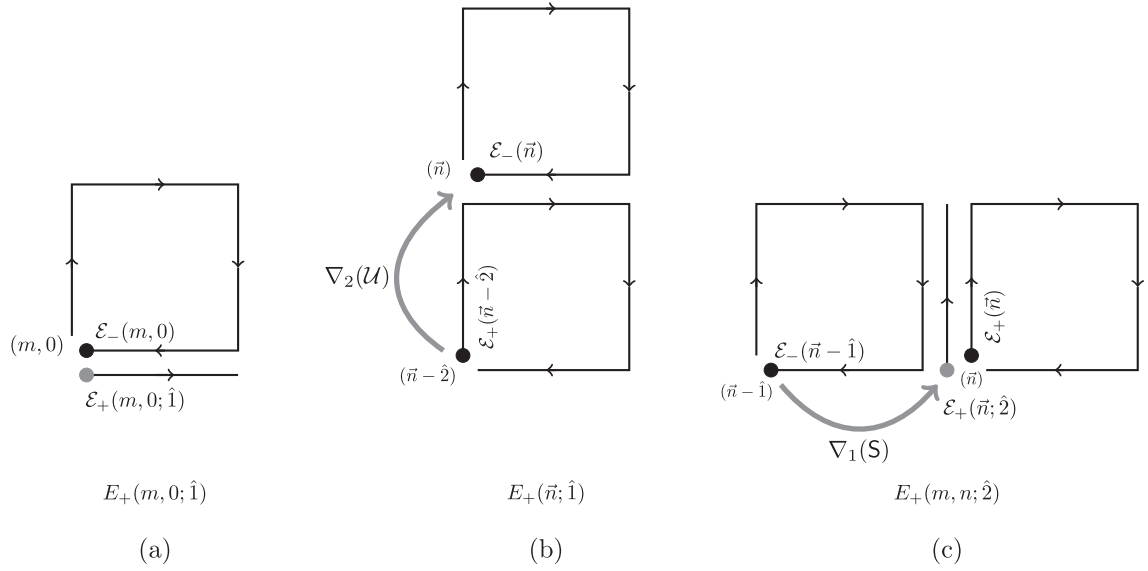

FIG. 8. Inverse canonical relations: Kogut-Susskind electric fields in terms of the dual plaquette and string electric fields. (a) Horizontal left electric field $E_+(m,0;\hat{1})$ in (26a). (b) Horizontal left electric fields $E_+(m,n;\hat{1})$, $n \neq 0$ in (26b). (c) Vertical left electric field $E_+(m,n;\hat{2})$ in (26c). The plaquette electric fields $\mathcal{E}(m,n)$ are shown by • and string electric fields $\mathcal{E}(m,n;\hat{i})$ are shown by •. The round arrows show that required parallel transports.

Like in Sec. III A 1, the conjugate electric fields are

$$\mathcal{E}_+(m,0;\hat{1}) = E_+(m,0;\hat{1}) - \sum_{j=1}^{N} \mathcal{S}_j(m,0) E_-(m+1,j;\hat{1}) \mathcal{S}_j^{-1}(m,0), \tag{32a}$$

$$\mathcal{E}_+(m,n;\hat{2}) = E_+(m,n;\hat{2}) - \sum_{j=n+1}^{N} \mathcal{S}'_j(m,n) E_-(m,j;\hat{1}) \mathcal{S}'^{-1}_j(m,n) + \sum_{j=n+1}^{N} \mathcal{S}_j(m,n) E_-(m+1,j;\hat{1}) \mathcal{S}_j^{-1}(m,n). \tag{32b}$$

As before, the vertical parallel transports are

$$\mathcal{S}'_j(m,n) = \prod_{k=n}^{j-1} U(m,k;\hat{2}) \tag{33}$$

and $\mathcal{S}'_0(m,0) = 1$. The canonical transformations ensure that all $N(N+1)$ vertical string pairs $(\mathcal{E}^a_+(\vec{n};\hat{2}), \mathcal{U}_{\alpha\beta}(\vec{n};\hat{2}))$ and $N$ horizontal string pairs $(\mathcal{E}^a_+(m,0;\hat{1}), \mathcal{U}_{\alpha\beta}(m,0;\hat{1}))$ satisfy the standard canonical commutation relations.

As before, under SU(N) gauge transformations (7) the plaquette conjugate pairs transform as adjoint matter (18) and the string conjugate pairs transform as gauge fields (23).

#### 2. Inverse relations

We can get Kogut-Susskind link operators from plaquette and string holonomies by solving Eqs. (28) and (31);

$$U(\vec{n};\hat{1}) = \mathsf{S}(\vec{n};\hat{1}), \tag{34a}$$

$$U(\vec{n};\hat{2}) = \mathcal{U}(\vec{n};\hat{2}). \tag{34b}$$

Where $\mathsf{S}(\vec{n};\hat{1})$ is the shortest path containing the plaquette holonomy $\mathcal{W}$, which connects the sites $\vec{n}$ and $\vec{n} + \hat{1}$ [see Fig. 9(a)]:

$$\mathsf{S}(\vec{n};\hat{1}) = \mathcal{L}(\vec{n}) \mathcal{U}(m,0;\hat{1}) \mathcal{R}(\vec{n}+\hat{1}). \tag{35}$$

In (35), we have defined left and right parallel transports

$$\mathcal{L}(\vec{n}) \equiv \left[\prod_{j=1}^{n} \mathcal{U}^\dagger(m,n-j;\hat{2}) \mathcal{W}(m,n-j)\right], \tag{36a}$$

$$\mathcal{R}(\vec{n}+\hat{1}) \equiv \left[\prod_{k=0}^{n-1} \mathcal{U}(m+1,k;\hat{2})\right]. \tag{36b}$$

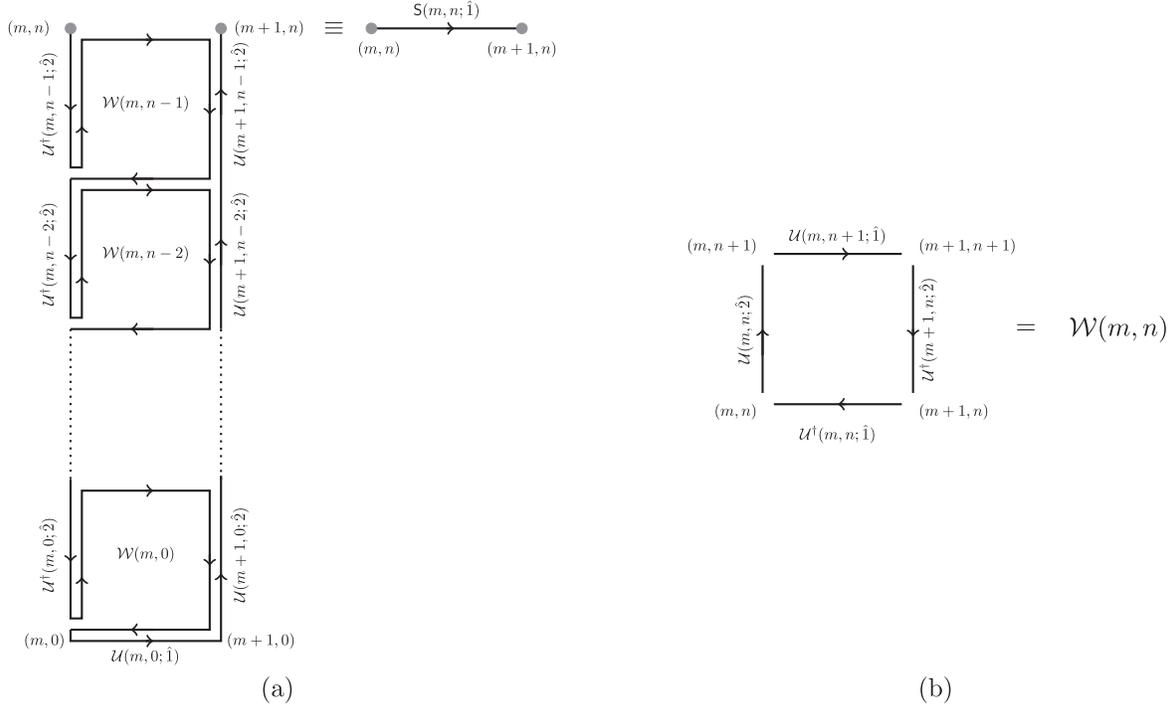

FIG. 9. (a) Nonlocal interactions in the horizontal direction between $\mathcal{E}_-(\vec{n}-\hat{1})$ and $\mathcal{E}_+(\vec{n})$ in (39). The parallel transport denoted by $\mathsf{S}(m,n;\hat{1})$ depends nonlocally on $\mathcal{U}$ and $\mathcal{W}$ leading to nonlocal interactions in (39). (b) Introduction of new gauge fields $\mathcal{U}(m,n;\hat{1})$ through the local plaquette constraints (41) converts them into minimal couplings.

The canonical relations (29), (32a), and (32b) can also be inverted to get the Kogut-Susskind electric fields in terms of the plaquette and string fields:

$$E_+(m,0;\hat{1}) = \mathcal{E}_-(m,0) + \mathcal{E}_+(m,0;\hat{1}), \quad (37a)$$

$$E_+(\vec{n};\hat{1}) = \nabla_2(\mathcal{U})\mathcal{E}(n), \quad (37b)$$

$$E_+(\vec{n};\hat{2}) = -\nabla_1(\mathcal{S})\mathcal{E}(\vec{n}) + \mathcal{E}_+(\vec{n};\hat{2}). \quad (37c)$$

In the inverse duality relations (37a), (37b), (37c) we have defined the difference operators with local $\mathcal{U}(\vec{n}-\hat{2};\hat{2})$ and nonlocal $\mathsf{S}^\dagger(\vec{n}-\hat{1};\hat{1})$ parallel transports as

$$\nabla_2(\mathcal{U})\mathcal{E}(\vec{n}) \equiv \mathcal{E}_-(\vec{n}) + \mathcal{U}^\dagger(\vec{n}-\hat{2};\hat{2})\mathcal{E}_+(\vec{n}-\hat{2})\mathcal{U}(\vec{n}-\hat{2};\hat{2}), \quad (38a)$$

$$\nabla_1(\mathcal{S})\mathcal{E}(\vec{n}) \equiv -\mathcal{E}_+(\vec{n}) - \mathsf{S}^\dagger(\vec{n}-\hat{1};\hat{1})\mathcal{E}_-(\vec{n}-\hat{1})\mathsf{S}(\vec{n}-\hat{1};\hat{1}). \quad (38b)$$

Note that after duality the Kogut electric fields are not fundamental. They are instead expressed in terms of electric scalar potentials $\mathcal{E}(m,n)$. These electric scalar potentials describe the gauge theory interactions in the dual version (39) or (49) as opposed to the magnetic vector potentials $U(m,n;\hat{i})$ that describe interactions in the original Hamiltonian (1). These dynamical issues are discussed in the next section.

## IV. SU(N) DUAL DYNAMICS

The Kogut-Susskind Hamiltonian (1) can now be rewritten in terms of the dual plaquette and string operators as [16]

$$H = \sum_{\vec{n}} \left[ g^2 \mathrm{Tr}\left( (\nabla_2(\mathcal{U})\mathcal{E}(\vec{n}))^2 + (\mathcal{E}_+(\vec{n},\hat{2}) - \nabla_1(\mathsf{S})\mathcal{E}(\vec{n}))^2 \right) + \frac{K}{g^2}(2N - \mathrm{Tr}(\mathcal{W}(\vec{n}) + \mathcal{W}^\dagger(\vec{n}))) \right]. \quad (39)$$

This dual or loop description is invariant under SU(N) gauge transformations (18), (23) and simple to interpret as follows: The original nontrivial four link interaction term in (1), which dominates near the $g^2 \to 0$ continuum limit, is now a simple noninteracting magnetic field term $\frac{1}{g^2}\mathrm{Tr}(\mathcal{W}+\mathcal{W}^\dagger) \sim \frac{1}{g^2}\vec{B}^2$. This is one of the expected outcomes of duality transformations. On the other hand, the original noninteracting electric field terms in (1) now describe the interactions in terms of the adjoint electric scalar potentials. Note that the dual interaction in the $y$ direction in (39) are the minimal coupling terms between electric scalar potentials and the string fields in the $y$ direction.

The immediate problem we face with the above dual description is the nonlocal and asymmetric dynamics due to the presence of $S(\vec{n}; \hat{1})$ in (37c). The underlying reason for this nonlocality and asymmetry is simply the absence of the horizontal holonomies that have been canonically transformed into $\mathcal{W}(m, n)$ as shown in Fig. 4(b). The asymmetric Gauss law constraints associated with the SU(N) gauge invariance (18) and (23) are

$$\mathcal{G}^a(\vec{n}) = \mathcal{E}_-^a(\vec{n}) + \mathcal{E}_+^a(\vec{n}) + \mathcal{E}_+^a(\vec{n}; \hat{2}) + \mathcal{E}_-^a(\vec{n}; \hat{2}) = 0. \quad (40)$$

The above constraints directly follow from the new configurations in Fig. 4(b). As shown in Appendix B the new Gauss law constraints (40) reduce to the old symmetric Gauss law constraints (8) when the canonical relations are used and thus confirming (29) and (32a), (32b). The next section addresses and solves the asymmetry and nonlocality issues by introducing new plaquette constraints.

### A. Plaquette constraints

Having obtained the dual magnetic field description in terms of the physical conjugate loop pairs $(\mathcal{E}(\vec{n}), \mathcal{W}(\vec{n}))$, we resolve the above asymmetry and nonlocality problems by reintroduction of horizontal link holonomies $\mathcal{U}(\vec{n}; \hat{1})$ through the local plaquette constraints:

$$\mathcal{U}(\vec{n}; \hat{2})\mathcal{U}(\vec{n} + \hat{2}; \hat{1})\mathcal{U}^\dagger(\vec{n} + \hat{1}; \hat{2})\mathcal{U}^\dagger(\vec{n}; \hat{1}) = \mathcal{W}(\vec{n}). \quad (41)$$

Note that the constraints (41) imposed on the dual theory are consistent with the dual gauge transformations (18) and (23). They physically mean that the newly created gauge invariant Wilson loops with gauge fields $(\mathcal{U}(\vec{n}; \hat{1}), \mathcal{U}(\vec{n}; \hat{2}))$ do not lead to any additional physical degrees of freedom. The motivation for introducing (41) is that on the constrained surface

$$\mathcal{U}(\vec{n}; \hat{1}) = S(\vec{n}; \hat{1}). \quad (42)$$

Now the nonlocal inverse relation (37c) takes the local form and we write

$$E_+(\vec{n}; \hat{i}) = \delta_{i2}\mathcal{E}_+(\vec{n}; \hat{i}) + \epsilon_{ij}\nabla_j(\mathcal{U})\mathcal{E}(\vec{n}). \quad (43)$$

In (43) $i, j = 1, 2$. The plaquette constraints (41) must commute with the Hamiltonian $H$ in (39). It is clear that the magnetic part, $H_M \sim \text{Tr}\,\mathcal{W}(\vec{n})$, commutes with (41) as $\mathcal{W}_{\alpha\beta}(\vec{n})$ and $\mathcal{U}_{\alpha\beta}(\vec{n}; \hat{i})$ are mutually independent and commuting dual degrees of freedom. It is easy to see that the constraints (41) will commute with the electric part $H_E$ ($H_E \sim \vec{E}^2(\vec{n}; \hat{1}) + \vec{E}^2(\vec{n}; \hat{2})$) also if the electric fields $E_+^a(\vec{n}; \hat{1})$ and $E_+^a(\vec{n}; \hat{2})$ defined by (37b) and (37c) rotate both sides of (41) covariantly. We therefore introduce electric fields $\mathcal{E}_+(\vec{n}; \hat{1})$, which are conjugate to auxiliary gauge fields $\mathcal{U}(\vec{n}; \hat{1})$ and write

$$E_+(\vec{n}; \hat{i}) = \mathcal{E}_+(\vec{n}; \hat{i}) + \epsilon_{ij}\nabla_j(\mathcal{U})\mathcal{E}(\vec{n}). \quad (44)$$

In (44) the covariant derivatives are defined as

$$\nabla_2(\mathcal{U})\mathcal{E}(\vec{n}) \equiv \mathcal{E}_-(\vec{n}) + \mathcal{U}^\dagger(\vec{n} - \hat{2}; \hat{2})\mathcal{E}_+(\vec{n} - \hat{2})\mathcal{U}(\vec{n} - \hat{2}; \hat{2}), \quad (45a)$$

$$\nabla_1(\mathcal{U})\mathcal{E}(\vec{n}) \equiv -\mathcal{E}_+(\vec{n}) - \mathcal{U}^\dagger(\vec{n} - \hat{1}; \hat{1})\mathcal{E}_-(\vec{n} - \hat{1})\mathcal{U}(\vec{n} - \hat{1}; \hat{1}). \quad (45b)$$

As mentioned before the parallel transports in (45a) and (45b) are also consistent with SU(N) gauge covariance. This provides an additional cross check for the validity of the SU(N) canonical or duality transformations.

At this stage it is interesting as well as illustrative to compare Eq. (44) with the corresponding equation in U(1) or Z(N) lattice gauge theories [9]. In U(1) case the Gauss law constraints in (2 + 1) dimension are

$$\vec{\nabla} \cdot \vec{E}(\vec{n}) \equiv \sum_{i=1}^{2}(\nabla_i E(\vec{n}; \hat{i})) = 0, \quad (46)$$

where $\nabla_i$ is the simple difference operator in $i = 1, 2$ directions. The obvious solutions defining the Abelian electric scalar potentials in (2 + 1) dimension are

$$E(\vec{n}; \hat{i}) = \epsilon_{ij}\nabla_j\mathcal{E}(\vec{n}). \quad (47)$$

The SU(N) electric scalar potentials defining Eq. (44) are obvious generalizations of the corresponding Abelian equation (47) with the ordinary difference operators replaced by the SU(N) covariant difference operators. Note that instead of directly solving (46) to obtain dual electric potential $\mathcal{E}(\vec{n})$ in (47), we can also use the present canonical transformation route to reach the same result. In U(1) case the parallel transports in (5) and (38a), (38b) are simple Abelian phase factors and cancel out. Thus there are no strings or link gauge fields and we recover (47) without any nonlocality or asymmetry problems.

The SU(N) Gauss law constraints

$$\mathcal{G}^a(\vec{n}) = \mathcal{E}_-^a(\vec{n}) + \mathcal{E}_+^a(\vec{n}) + \sum_{i=1,2}(\mathcal{E}_-^a(\vec{n}; \hat{i}) + \mathcal{E}_+^a(\vec{n}; \hat{i})) = 0 \quad (48)$$

are now symmetric as shown in Fig. 10. Under SU(N) gauge transformations all electric fields appearing in (48) transform like adjoint matter fields.

The new Hamiltonian that commutes with the constraints (41) written in terms of the dual operators is

$$H = \sum_{\vec{n}} \left[ g^2 \operatorname{Tr} \sum_{i=1}^{2} \left( \mathcal{E}_+(\vec{n},\hat{i}) + \epsilon_{ij}\nabla_j(\mathcal{U})\mathcal{E}(\vec{n}) \right)^2 + \frac{K}{g^2}(2N - \operatorname{Tr}(\mathcal{W}(\vec{n}) + \mathcal{W}^\dagger(\vec{n}))) \right]. \quad (49)$$

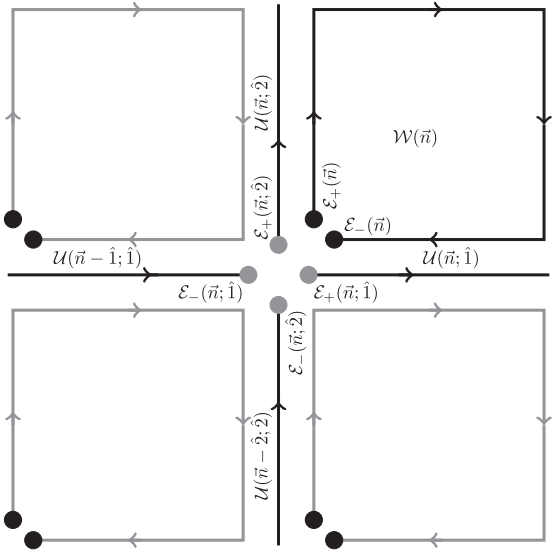

FIG. 10. The new symmetric Gauss Laws $\mathcal{E}_-^a(\vec{n}) + \mathcal{E}_+^a(\vec{n}) + \sum_{i=1}^{2}(\mathcal{E}_+^a(\vec{n};\hat{i}) + \mathcal{E}_-^a(\vec{n};\hat{i})) = 0$. There are six electric fields at each site $\vec{n}$. The two plaquette electric fields or electric scalar potentials are shown by dark bullets • and the four string electric fields are shown by gray bullets •.

The dual Hamiltonian (49) can also be interpreted as the loop Hamiltonian. Its physical interpretation is very simple. The second interacting term in (1) dualizes to the noninteracting magnetic field term in (49). It creates and annihilates the single plaquette loops. This is most transparent in the prepotential operator language [9,13]. The first original noninteracting electric field term in (1) dualizes to the loop-loop interaction term in (49). These SU(N) loop interactions are through minimal couplings of the loop electric scalar potential to the gauge fields. This duality between interacting and noninteracting terms leads to inversion of the coupling constant: $g^2 \rightarrow \frac{1}{g^2}$. Note that the physical degrees of freedom are associated only with the SU(N) magnetic fields and their conjugate electric potentials $(\mathcal{E}(\vec{n}), \mathcal{W}(\vec{n}))$. The auxiliary string sector $(\mathcal{E}(\vec{n};\hat{i}), \mathcal{U}(m,\vec{n};\hat{i}))$ with the new constraints (41) makes the dual description local as well as simple and rotationally covariant.

We again emphasize that the $N^2$ horizontal strings $\mathcal{U}(m,n>0;\hat{1})$ can be removed using the $N^2$ constraints (41). As a result their $N^2$ conjugate electric fields $\mathcal{E}(m,n>0;\hat{1})$ can be put equal to zero without loss of any generality. We thus recover the nonlocal Hamiltonian (39), which in turn is exactly equivalent to the Kogut-Susskind Hamiltonian (1) due to the canonical transformations. In fact, at this stage we can also remove the vertical strings completely. Such SU(N) canonical or duality transformations leading to dual SU(N) spin model without any gauge or string degrees of freedom have been studied in the past [9]. They lead to nonlocal dynamics. In the present framework, with all interactions local and proportional to $g^2$, the dual Hamiltonian see (49) can be used to set up a weak coupling perturbation theory near the continuum $g^2 \rightarrow 0$ limit. The matter fields can be coupled to the SU(N) gauge fields $\mathcal{U}(\vec{n};\hat{i})$ through minimal coupling so that the SU(N) gauge invariance (18) and (23) remains intact.

## V. SUMMARY AND DISCUSSION

In this work we have constructed the canonical transformations in SU(N) lattice gauge theory that lead to local dual Hamiltonian with minimal interactions between dual electric scalar potentials and the auxiliary gauge fields. This result is easy to understand as under gauge transformations the magnetic or plaquette loop fields as well as their conjugate electric scalar potentials transform like SU(N) adjoint matter fields. The transformations convert the plaquette interaction terms into the pure noninteracting magnetic field terms and the pure noninteracting electric field terms into the electric scalar potential minimal coupling interaction terms. These results are important as the plaquette interaction terms involving four links, which dominate near the continuum $g^2 \rightarrow 0$ limit, have been completely simplified. In the past, even in the simple SU(2) lattice gauge theory case, these plaquette interactions become extremely complicated in the loop Hilbert space [5,13]. Therefore, it will be interesting to develop a systematic weak coupling loop perturbation theory in the $g^2 \rightarrow 0$ continuum limit with the dual Hamiltonian (49).

In the context of quantum simulations of non-Abelian lattice gauge theories [6–8], the problem with the electric basis [7,13], is the complicated matrix elements of the magnetic field terms $H_M$ in (1). On the other hand, in the magnetic basis the electric field terms $H_E$ in (1) becomes complicated and nonlocal [6,8,9]. The present work provides a magnetic basis without the above nonlocality problem and therefore may be better suited for quantum simulations near the continuum limit.

In $(3 + 1)$ dimension these canonical transformation can be carried out on every $(XZ)$ and $(YZ)$ plane similar to the present $(2 + 1)$ dimensional case. We thus convert all $X$, $Y$ links at $z > 0$ into $(XZ)$, $(YZ)$ plaquettes, respectively, and $Z$ links into the unphysical strings. Now the dual formulation will have nonlocality in both the electric and magnetic field parts of the Hamiltonian. The nonlocality in the electric field part, like in $(2 + 1)$ dimension, will be due to gauge invariance, whereas the absence of $(XY)$ plaquette will introduce nonlocality in the magnetic part. This nonlocal dynamics can again be made local by introducing new plaquette constraints. The work in these directions is in progress and will be reported elsewhere.

## APPENDIX A: CANONICAL TRANSFORMATION ON A 2 × 2 PLAQUETTE LATTICE

In this appendix, we will explicitly work out canonical transformations (14), (20a), and (20b) for the simple $2 \times 2$ plaquette lattice. Starting from the top-left plaquette, we make canonical transformations over four plaquettes in the following four steps I, II, III and IV to construct the plaquettes $\mathcal{W}(0,1), \mathcal{W}(0,0), \mathcal{W}(1,1)$, and $\mathcal{W}(1,0)$, respectively (see Fig. 11). Each of these four steps involves three gluings of Kogut-Susskind link holonomies through canonical transformations illustrated in Fig. 3.

### 1. Construction of $\mathcal{W}(0,1)$

In the first step we glue four links of the top-left plaquette in the clockwise direction and convert them into the plaquette $\mathcal{W}(0,1)$ and the three remaining holonomies: $\mathcal{U}(0,1;\hat{2}), \tilde{\mathcal{U}}(1,1;\hat{2}), \tilde{\mathcal{U}}(0,1;\hat{1})$ as shown in Fig. 12(I). The three canonical transformations involved in this first step are as follows:

(i) The first canonical transformation is

$$\begin{bmatrix} (E_+(0,1;\hat{2}), U(0,1;\hat{2})) \\ (E_+(0,2;\hat{1}), U(0,2;\hat{1})) \end{bmatrix}$$
$$\rightarrow \begin{bmatrix} (\mathcal{E}_+(0,1;\hat{2}), \mathcal{U}(0,1;\hat{2})) \\ (\tilde{\mathcal{E}}_+(0,1), \tilde{\mathcal{U}}(0,1)) \end{bmatrix}.$$

The two new holonomies are defined as

$$\mathcal{U}(0,1;\hat{2}) = U(0,1;\hat{2}), \qquad (A1)$$

$$\tilde{\mathcal{U}}(0,1) = U(0,1;\hat{2})U(0,2;\hat{1}). \qquad (A2)$$

The basic canonical transformations (9) determine their right electric fields

$$\mathcal{E}_-(0,2;\hat{2}) = E_-(0,2;\hat{2}) + E_+(0,2;\hat{1}) \qquad (A3)$$

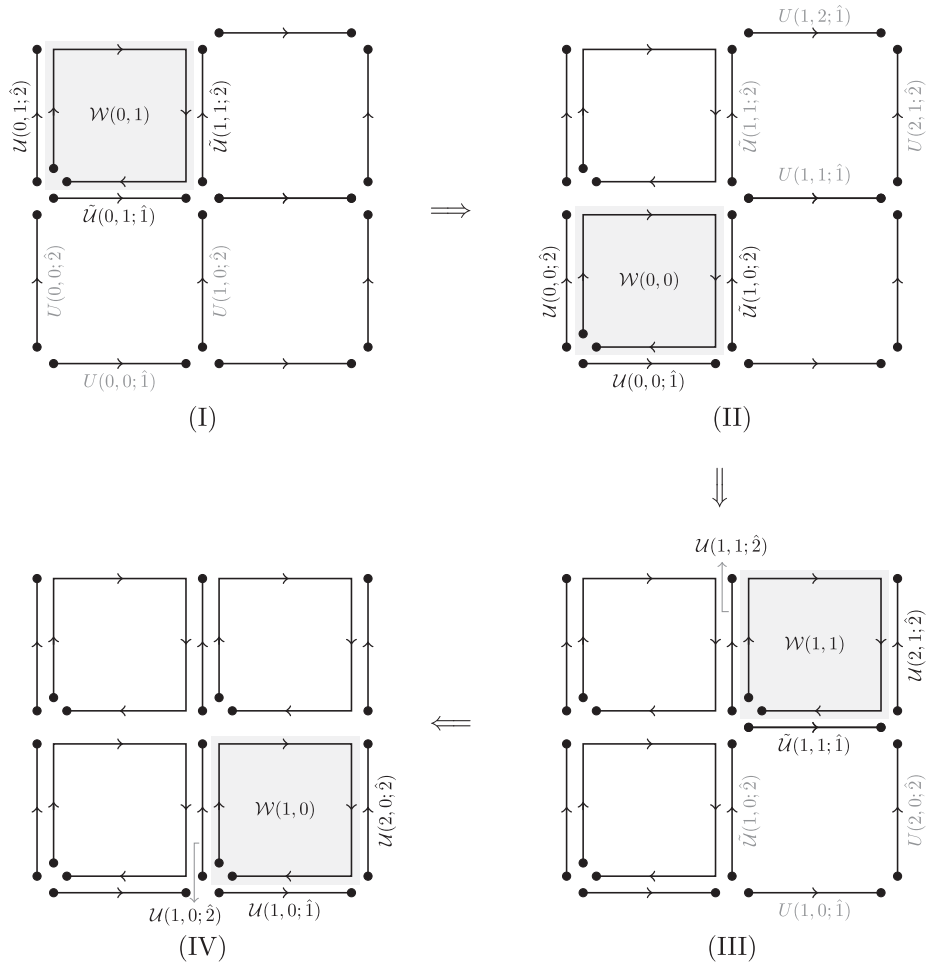

FIG. 11. Four steps canonical transformations on a simple $2 \times 2$ plaquette lattice. We construct $\mathcal{W}(0,1), \mathcal{W}(0,0), \mathcal{W}(1,1)$, and $\mathcal{W}(1,0)$ sequentially in steps I, II, III, and IV.

$$\tilde{\mathcal{E}}_-(0,1) = E_-(1,2;\hat{1}). \tag{A4}$$

The canonical transformations (A1), (A2), (A3), (A4) and the electric field locations are shown in Fig. 12(a). Their corresponding left electric fields can be obtained by parallel transports as in (5):

$$\mathcal{E}_+(0,1;\hat{2}) = E_+(0,1;\hat{2}) + \mathcal{S}_2(0,1) \\ \times E_-(1,2;\hat{1})\mathcal{S}_2^{-1}(0,1), \tag{A5}$$

$$\tilde{\mathcal{E}}_+(0,1) = -\tilde{\mathcal{U}}(0,1)E_-(1,2;\hat{1})\tilde{\mathcal{U}}^\dagger(0,1). \tag{A6}$$

In (A5) we have identified

$$\mathcal{S}_{j=2}(0,1) \equiv \tilde{\mathcal{U}}(0,1) = U(0,1;\hat{2})U(0,2;\hat{1}). \tag{A7}$$

We thus obtain (20b) at $m=0$, $n=1$.
(ii) The second canonical transformation is

$$\begin{bmatrix} (\tilde{\mathcal{E}}_+(0,1), \tilde{\mathcal{U}}(0,1)) \\ (E_+(1,1;\hat{2}), U(1,1;\hat{2})) \end{bmatrix} \\ \rightarrow \begin{bmatrix} (\tilde{\tilde{\mathcal{E}}}_+(0,1), \tilde{\tilde{\mathcal{U}}}(0,1)) \\ (\tilde{\mathcal{E}}_+(1,1;\hat{2}), \tilde{\mathcal{U}}(1,1;\hat{2})) \end{bmatrix}.$$

The two new holonomies are defined as

$$\tilde{\mathcal{U}}(1,1;\hat{2}) = U(1,1;\hat{2}), \tag{A8}$$

$$\tilde{\tilde{\mathcal{U}}}(0,1) = \tilde{\mathcal{U}}(0,1)U^\dagger(1,1;\hat{2}), \\ = U(0,1;\hat{2})U(0,2;\hat{1})U^\dagger(1,1;\hat{2}). \tag{A9}$$

The canonical transformations (9) lead to the following electric fields

$$\tilde{\mathcal{E}}_-(1,2;\hat{2}) = E_-(1,2;\hat{2}) + \tilde{\mathcal{E}}_-(0,1), \\ = E_-(1,2;\hat{2}) + E_-(1,2;\hat{1}), \tag{A10}$$

$$\tilde{\tilde{\mathcal{E}}}_+(0,1) = \tilde{\mathcal{E}}_+(0,1), \\ = -\mathcal{S}_2(0,1)E_-(1,2;\hat{1})\mathcal{S}_2^{-1}(0,1). \tag{A11}$$

The canonical transformations (A8), (A9), (A10), (A11) and the electric field locations are shown in Fig. 12(b). We now using (5), we obtain the left and right electric fields of $\tilde{\mathcal{U}}(1,1;\hat{2})$ and $\tilde{\tilde{\mathcal{U}}}(0,1)$, respectively, for later use:

$$\tilde{\mathcal{E}}_+(1,1;\hat{2}) = E_+(1,1;\hat{2}) - \mathcal{S}'_2(1,1) \\ \times E_-(1,2;\hat{1})\mathcal{S}'^{-1}_2(1,1), \tag{A12}$$

$$\tilde{\tilde{\mathcal{E}}}_-(0,1) = \mathcal{S}'_2(1,1)E_-(1,2;\hat{1})\mathcal{S}'^{-1}_2(1,1). \tag{A13}$$

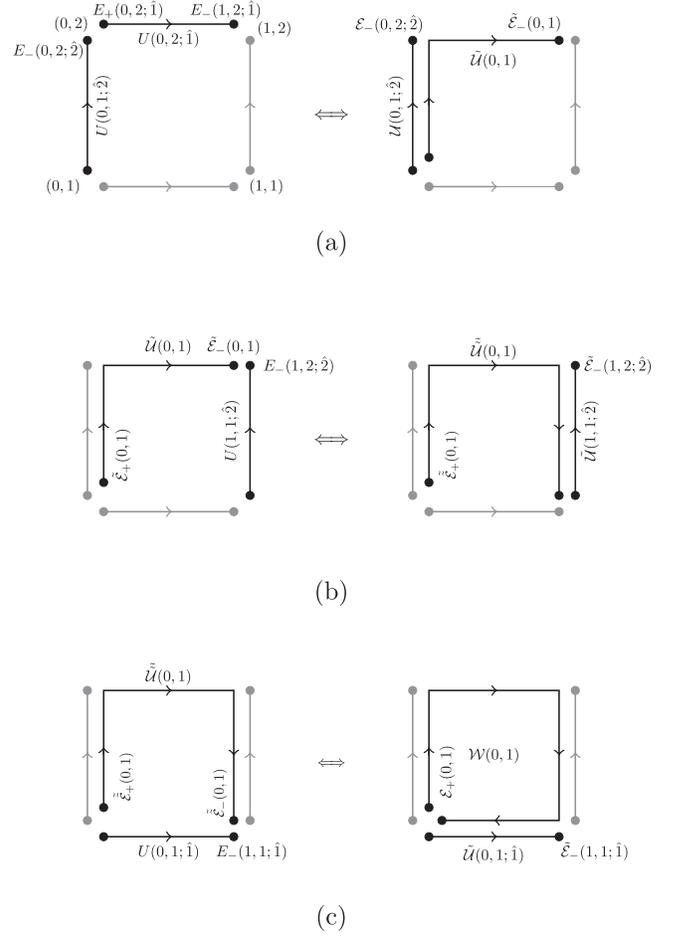

(a)

(b)

(c)

FIG. 12. First step of canonical transformation over the top leftmost plaquette for $2\times 2$ lattice. This step is further divided into three parts (a), (b), and (c).

In (A13) we have identified

$$\mathcal{S}'_2(1,1) \equiv U(1,1;\hat{2}) \tag{A14}$$

as defined in the (21c) for $m=1$.
(iii) The third canonical transformation is

$$\begin{bmatrix} (\tilde{\tilde{\mathcal{E}}}_+(0,1), \tilde{\tilde{\mathcal{U}}}(0,1)) \\ (E_+(0,1;\hat{1}), U(0,1;\hat{1})) \end{bmatrix} \\ \rightarrow \begin{bmatrix} (\mathcal{E}_+(0,1), \mathcal{U}(0,1)) \\ (\tilde{\mathcal{E}}_+(0,1;\hat{1}), \tilde{\mathcal{U}}(0,1;\hat{1})) \end{bmatrix}.$$

We now glue the conjugate pair $(\tilde{\tilde{\mathcal{E}}}_+(0,1), \tilde{\tilde{\mathcal{U}}}(0,1))$ obtained in the previous step with link conjugate link pair $(E_+(0,1;\hat{1}), U(0,1;\hat{1}))$ to get the first plaquette conjugate pair $(\mathcal{E}_+(0,1), \mathcal{W}(0,1))$ along with the intermediate link conjugate pair $(\tilde{\mathcal{E}}_+(0,1;\hat{1}), \tilde{\mathcal{U}}(0,1;\hat{1}))$

$$\tilde{\mathcal{U}}(0,1;\hat{1}) = U(0,1;\hat{1}), \quad (A15)$$

$$\begin{aligned}\mathcal{W}(0,1) &= \tilde{\tilde{\mathcal{U}}}(0,1)U^{\dagger}(0,1;\hat{1}), \\ &= U(0,1;\hat{2})U(0,2;\hat{1})U^{\dagger}(1,1;\hat{2})U^{\dagger}(0,1;\hat{1}).\end{aligned} \quad (A16)$$

Their conjugate electric fields are

$$\begin{aligned}\tilde{\mathcal{E}}_{-}(1,1;\hat{1}) &= E_{-}(1,1;\hat{1}) + \tilde{\tilde{\mathcal{E}}}_{-}(0,1), \\ &= E_{-}(1,1;\hat{1}) + \mathcal{S}'_2(1,1) \\ &\quad \times E_{-}(1,2;\hat{1})\mathcal{S}'^{-1}_2(1,1),\end{aligned} \quad (A17)$$

$$\begin{aligned}\mathcal{E}_{+}(0,1) &= \tilde{\tilde{\mathcal{E}}}_{+}(0,1), \\ &= -\mathcal{S}_2(0,1)E_{-}(1,2;\hat{1})\mathcal{S}^{-1}_2(0,1).\end{aligned} \quad (A18)$$

Above canonical transformations are shown in Fig. 12(c). We thus obtain (14) for $m = 0$, $n = 1$.

The above three canonical transformations complete step I. In summary, starting from the four link holonomies,

$$(U(0,1;\hat{2}), U(0,2;\hat{1}), U(1,1;\hat{2}), U(0,1;\hat{1})),$$

we have obtained the following four equivalent holonomies

$$\begin{aligned}\mathcal{U}(0,1;\hat{2}) &= U(0,1;\hat{2}), \\ \mathcal{W}(0,1) &= U(0,1;\hat{2})U(0,2;\hat{1})U^{\dagger}(1,1;\hat{2})U^{\dagger}(0,1;\hat{1}), \\ \tilde{\mathcal{U}}(1,1;\hat{2}) &= U(1,1;\hat{2}), \\ \tilde{\mathcal{U}}(0,1;\hat{1}) &= U(0,1;\hat{1}).\end{aligned} \quad (A19)$$

The corresponding electric fields are

$$\begin{aligned}\mathcal{E}_{+}(0,1;\hat{2}) &= E_{+}(0,1;\hat{2}) + \mathcal{S}_2(0,1)E_{-}(1,2;\hat{1})\mathcal{S}^{-1}_2(0,1), \\ \mathcal{E}_{+}(0,1) &= -\mathcal{S}_2(0,1)E_{-}(1,2;\hat{1})\mathcal{S}^{-1}_2(0,1), \\ \tilde{\mathcal{E}}_{-}(1,2;\hat{2}) &= E_{-}(1,2;\hat{2}) + E_{-}(1,2;\hat{1}), \\ \tilde{\mathcal{E}}_{-}(1,1;\hat{1}) &= E_{-}(1,1;\hat{1}) + \mathcal{S}'_2(1,1)E_{-}(1,2;\hat{1})\mathcal{S}'^{-1}_2(1,1).\end{aligned} \quad (A20)$$

Now we notice that in step I, we have traded off $U(0,2;\hat{1})$ into the plaquette $\mathcal{W}(0,1)$ so its electric field $E_{-}(1,2;\hat{1})$ appears in all the four new plaquette and string electric fields with appropriate parallel transports (A7) and (A14). Now we will perform steps II, III, and IV using Eqs. (A19) and (A20).

So we have two dual holonomies as required and two intermediate holonomies that will be used for canonical transformation in steps II and III. Electric fields for these holonomies are, see Fig. 12.

### 2. Construction of $\mathcal{W}(0,0)$

In the second step, we consider four holonomies $U(0,0;\hat{2})$, $\tilde{\mathcal{U}}(0,1;\hat{1})$, $U(1,0;\hat{2})$, and $U(0,0;\hat{1})$, see Fig. 11(I), and canonically convert them into following four holonomies, see Fig. 11(II):

$$\begin{aligned}\mathcal{U}(0,0;\hat{2}) &= U(0,0;\hat{2}), \\ \mathcal{W}(0,0) &= U(0,0;\hat{2})\tilde{\mathcal{U}}(0,1;\hat{1})U^{\dagger}(1,0;\hat{2})U^{\dagger}(0,0;\hat{1}), \\ &= U(0,0;\hat{2})U(0,1;\hat{1})U^{\dagger}(1,0;\hat{2})U^{\dagger}(0,0;\hat{1}), \\ \tilde{\mathcal{U}}(1,0;\hat{2}) &= U(1,0;\hat{2}), \\ \mathcal{U}(0,0;\hat{1}) &= U(0,0;\hat{1}),\end{aligned} \quad (A21)$$

with their electric fields given by

$$\begin{aligned}\mathcal{E}_{+}(0,0;\hat{2}) &= E_{+}(0,0;\hat{2}) + \mathcal{S}_1(0,0)\tilde{\mathcal{E}}_{-}(1,1;\hat{1})\mathcal{S}^{-1}_1(0,0), \\ \mathcal{E}_{+}(0,0) &= -\mathcal{S}_1(0,0)\tilde{\mathcal{E}}_{-}(1,1;\hat{1})\mathcal{S}^{-1}_1(0,0), \\ \tilde{\mathcal{E}}_{-}(1,1;\hat{2}) &= E_{-}(1,1;\hat{2}) + \tilde{\mathcal{E}}_{-}(1,1;\hat{1}), \\ \mathcal{E}_{-}(1,0;\hat{1}) &= E_{-}(1,0;\hat{1}) + \mathcal{S}'_1(1,0)\tilde{\mathcal{E}}_{-}(1,1;\hat{1})\mathcal{S}'^{-1}_1(1,0).\end{aligned} \quad (A22)$$

In Eq. (A22) we have identified strings

$$\mathcal{S}_1(0,0) \equiv U(0,0;\hat{2})U(0,1;\hat{1}), \quad (A23)$$

$$\mathcal{S}'_1(1,0) \equiv U(1,0;\hat{2}) \quad (A24)$$

as defined in (15a) for $m = 0$ and (21a) for $m = 0$, respectively. Now we can use the expression of $\tilde{\mathcal{E}}_{-}(1,1;\hat{1})$ given in step I to get the following:

$$\mathcal{E}_{+}(0,0;\hat{2}) = E_{+}(0,0;\hat{2}) + \sum_{j=1}^{2}\mathcal{S}_j(0,0)E_{-}(1,j;\hat{1})\mathcal{S}^{-1}_j(0,0), \quad (A25)$$

$$\mathcal{E}_+(0,0) = -\sum_{j=1}^{2} \mathcal{S}_j(0,0) E_-(1,j;\hat{1}) \mathcal{S}_j^{-1}(0,0), \quad (A26)$$

$$\tilde{\mathcal{E}}_-(1,1;\hat{2}) = E_-(1,1;\hat{2}) + E_-(1,1;\hat{1})$$
$$+ \mathcal{S}'_2(1,0) E_-(1,2;\hat{1}) \mathcal{S}'^{-1}_2(1,0), \quad (A27)$$

$$\mathcal{E}_-(1,0;\hat{1}) = E_-(1,0;\hat{1})$$
$$+ \sum_{j=1}^{2} \mathcal{S}'_j(1,0) E_-(1,j;\hat{1}) \mathcal{S}'^{-1}_j(1,0). \quad (A28)$$

In the above equations, we have identified strings

$$\mathcal{S}_2(0,0) \equiv U(0,0;\hat{2}) U(0,1;\hat{1}) U(1,1;\hat{2}), \quad (A29)$$

$$\mathcal{S}'_2(1,0) \equiv U(1,0;\hat{2}) U(1,1;\hat{2}) \quad (A30)$$

as defined in (15b) for $m = 0$ and (21b) for $m = 1$, respectively. Thus we have obtain (20b) and (14) for $m$, $n = 0$. We use (A27) and (A28) to obtain left electric fields for holonomies $\tilde{\mathcal{U}}(1,0;\hat{2})$ and $\mathcal{U}(0,0;\hat{1})$, respectively;

$$\tilde{\mathcal{E}}_+(1,0;\hat{2}) = E_+(1,0;\hat{2}) - \sum_{j=1}^{2} \mathcal{S}'_j(1,0) E_-(1,j;\hat{1}) \mathcal{S}'^{-1}_j(1,0), \quad (A31)$$

$$\mathcal{E}_+(0,0;\hat{1}) = E_+(0,0;\hat{1}) - \sum_{j=1}^{2} \mathcal{S}_j(0,0) E_-(1,j;\hat{1}) \mathcal{S}_j^{-1}(0,0). \quad (A32)$$

Equation (A31) will be used in step IV and (A32) is (20a) for $m = 0$.

### 3. Construction of $\mathcal{W}(1,1)$

In the third step, we start with four holonomies $\tilde{\mathcal{U}}(1,1;\hat{2})$, $U(1,2;\hat{1})$, $U(2,1;\hat{2})$, and $U(1,1;\hat{1})$ of top right plaquette, see Fig. 11(II), and performing canonical transformations similar to previous step II we convert them into one plaquette, two strings, and one intermediary holonomy:

$$\mathcal{U}(1,1;\hat{2}) = \tilde{\mathcal{U}}(1,1;\hat{2}) = U(1,1;\hat{2}),$$
$$\mathcal{W}(1,1) = \tilde{\mathcal{U}}(1,1;\hat{2}) U(1,2;\hat{1}) U^\dagger(2,1;\hat{2}) U^\dagger(1,1;\hat{1}),$$
$$= U(1,1;\hat{2}) U(1,2;\hat{1}) U^\dagger(2,1;\hat{2}) U^\dagger(1,1;\hat{1}),$$
$$\mathcal{U}(2,1;\hat{2}) = U(2,1;\hat{2}),$$
$$\tilde{\mathcal{U}}(1,1;\hat{1}) = U(1,1;\hat{1}). \quad (A33)$$

The above holonomies are shown in Fig. 11(III) and their electric fields are given by

$$\mathcal{E}_+(1,1;\hat{2}) = \tilde{\mathcal{E}}_+(1,1;\hat{2}) + \mathcal{S}_2(1,1) E_-(2,2;\hat{1}) \mathcal{S}_2^{-1}(1,1), \quad (A34)$$

$$\mathcal{E}_+(1,1) = -\mathcal{S}_2(1,1) E_-(2,2;\hat{1}) \mathcal{S}_2^{-1}(1,1), \quad (A35)$$

$$\mathcal{E}_-(2,2;\hat{2}) = E_-(2,2;\hat{2}) + E_-(2,2;\hat{1}), \quad (A36)$$

$$\tilde{\mathcal{E}}_-(2,1;\hat{1}) = E_-(2,1;\hat{1}) + \mathcal{S}'_2(2,1) E_-(2,2;\hat{1}) \mathcal{S}'^{-1}_2(2,1). \quad (A37)$$

In the above equations, we have identified strings

$$\mathcal{S}_2(1,1) \equiv U(1,1;\hat{2}) U(1,2;\hat{1}), \quad (A38)$$

$$\mathcal{S}'_2(2,1) \equiv U(2,1;\hat{2}) \quad (A39)$$

as defined in (15c) for $m = 1$ and (21c) for $m = 2$, respectively. Using (A12) into (A34) we obtain (20b) for $m = n = 1$ and (A35) is nothing but (14) for $m = n = 1$. Equation (A36) is used to write the left electric field of string $\mathcal{U}(2,1;\hat{2})$:

$$\mathcal{E}_+(2,1;\hat{2}) = E_+(2,1;\hat{2}) - \mathcal{S}'_2(2,1) E_-(2,2;\hat{1}) \mathcal{S}'^{-1}_2(2,1), \quad (A40)$$

which is (20b) for $m = 2$, $n = 1$, and (A37) is used for canonical transformations in step IV.

### 4. Construction of $\mathcal{W}(1,0)$

In the fourth step we take four holonomies $\tilde{\mathcal{U}}(1,0;\hat{2})$, $\tilde{\mathcal{U}}(1,1;\hat{1})$, $U(2,0;\hat{2})$, and $U(1,0;\hat{1})$, see Fig. 11(III), and canonically convert them into following four holonomies, see Fig. 11(IV):

$$\mathcal{U}(1,0;\hat{2}) = \tilde{\mathcal{U}}(1,0;\hat{2}) = U(1,0;\hat{2}),$$
$$\mathcal{W}(1,0) = U(1,0;\hat{2}) \tilde{\mathcal{U}}(1,1;\hat{1}) U^\dagger(2,0;\hat{2}) U^\dagger(1,0;\hat{1}),$$
$$= U(1,0;\hat{2}) U(1,1;\hat{1}) U^\dagger(2,0;\hat{2}) U^\dagger(1,0;\hat{1}),$$
$$\mathcal{U}(2,0;\hat{2}) = U(2,0;\hat{2}),$$
$$\mathcal{U}(1,0;\hat{1}) = U(1,0;\hat{1}), \quad (A41)$$

with their electric fields

$$\mathcal{E}_+(1,0;\hat{2}) = \tilde{\mathcal{E}}_+(1,0;\hat{2}) + \mathcal{S}_1(1,0) \tilde{\mathcal{E}}_-(2,1;\hat{1}) \mathcal{S}_1^{-1}(1,0),$$
$$\mathcal{E}_+(1,0) = -\mathcal{S}_1(1,0) \tilde{\mathcal{E}}_-(2,1;\hat{1}) \mathcal{S}_1^{-1}(1,0),$$
$$\mathcal{E}_-(2,1;\hat{2}) = E_-(2,1;\hat{2}) + \tilde{\mathcal{E}}_-(2,1;\hat{1}),$$
$$\mathcal{E}_-(2,0;\hat{1}) = E_-(2,0;\hat{1}) + \mathcal{S}'_1(2,0) \tilde{\mathcal{E}}_-(2,1;\hat{1}) \mathcal{S}'^{-1}_1(2,0). \quad (A42)$$

In the above equations, we have identified strings

$$\mathcal{S}_1(1,0) \equiv U(1,0;\hat{2})U(1,1;\hat{1}), \quad (A43)$$

$$\mathcal{S}'_1(2,0) \equiv U(2,0;\hat{2}) \quad (A44)$$

as defined in (15a) for $m=1$ and (21a) for $m=2$, respectively. Now we can use the expression of $\tilde{\mathcal{E}}_+(1,0;\hat{2})$ and $\tilde{\mathcal{E}}_-(2,1;\hat{1})$ obtained in steps II and III, respectively;

$$\mathcal{E}_+(1,0;\hat{2}) = E_+(1,0;\hat{2}) - \sum_{j=1}^{2} \mathcal{S}'_j(1,0) E_-(1,j;\hat{1}) \mathcal{S}'^{-1}_j(1,0)$$
$$+ \sum_{j=1}^{2} \mathcal{S}_j(0,0) E_-(2,j;\hat{1}) \mathcal{S}^{-1}_j(0,0), \quad (A45)$$

$$\mathcal{E}_+(1,0) = -\sum_{j=1}^{2} \mathcal{S}_j(1,0) E_-(2,j;\hat{1}) \mathcal{S}^{-1}_j(1,0), \quad (A46)$$

$$\mathcal{E}_-(2,1;\hat{2}) = E_-(2,1;\hat{2}) + E_-(2,1;\hat{1})$$
$$+ \mathcal{S}'_2(2,0) E_-(2,2;\hat{1}) \mathcal{S}'^{-1}_2(2,0), \quad (A47)$$

$$\mathcal{E}_-(2,0;\hat{1}) = E_-(2,0;\hat{1})$$
$$+ \sum_{j=1}^{2} \mathcal{S}'_j(2,0) E_-(2,j;\hat{1}) \mathcal{S}'^{-1}_j(2,0). \quad (A48)$$

In the above equations, we have identified strings

$$\mathcal{S}_2(1,0) \equiv U(1,0;\hat{2})U(1,1;\hat{1})U(2,1;\hat{2}), \quad (A49)$$

$$\mathcal{S}'_2(2,0) \equiv U(2,0;\hat{2})U(2,1;\hat{2}) \quad (A50)$$

as defined in (15b) for $m=0$ and (21b) for $m=1$, respectively. Equations (A45) and (A46) are (20b) and (14) for $m=1, n=0$, respectively. We use (A47) and (A48) to write left electric fields for holonomies $U(2,0;\hat{2})$ and $U(1,0;\hat{1})$, respectively.

$$\mathcal{E}_+(2,0;\hat{2}) = E_+(2,0;\hat{2}) - \sum_{j=1}^{2} \mathcal{S}'_j(2,0) E_-(2,j;\hat{1}) \mathcal{S}'^{-1}_j(2,0), \quad (A51)$$

$$\mathcal{E}_+(1,0;\hat{1}) = E_+(1,0;\hat{1}) - \sum_{j=1}^{2} \mathcal{S}_j(1,0) E_-(2,j;\hat{1}) \mathcal{S}^{-1}_j(1,0). \quad (A52)$$

Equations (A51) and (A52) are (20b) for $m=2, n=0$ and (20a) for $m=1, n=0$, respectively.

## APPENDIX B: CANONICAL TRANSFORMATIONS, DUALITY, AND GAUSS LAWS

In this appendix we show that the SU(N) Gauss laws in terms of the plaquette and string electric fields (40) reduce to the original Gauss laws (8) when the canonical transformations are used. This equivalence requires numerous highly nontrivial cancellations all along the nonlocal paths. Thus these calculations also validate the SU(N) canonical or duality transformations discussed in this work. The left plaquette electric fields [see (29)] are

$$\mathcal{E}_+(\vec{n}) = -\sum_{j=n+1}^{N} \mathcal{S}_j(\vec{n}) E_-(m+1,j;\hat{1}) \mathcal{S}^{-1}_j(\vec{n}). \quad (B1)$$

In (B1) $(\vec{n}) = (m,n)$. We can obtain the right electric field for plaquette operators by parallel transport

$$\mathcal{E}_-(\vec{n}) = -\mathcal{W}^\dagger(\vec{n}) \mathcal{E}_+(\vec{n}) \mathcal{W}(\vec{n}),$$
$$= U(\vec{n};\hat{1}) \sum_{j=n+1}^{N} \mathcal{S}'_j(m+1,n) E_-(m+1,j;\hat{1})$$
$$\mathcal{S}'^{-1}_j(m+1,n) U^\dagger(\vec{n};\hat{1}). \quad (B2)$$

The left electric field for vertical strings (31) are

$$\mathcal{E}_+(\vec{n};\hat{2}) = E_+(\vec{n};\hat{2}) - \sum_{j=n+1}^{N} \mathcal{S}'_j(\vec{n}) E_-(m,j;\hat{1}) \mathcal{S}'^{-1}_j(\vec{n})$$
$$+ \sum_{j=n+1}^{N} \mathcal{S}_j(\vec{n}) E_-(m+1,j;\hat{1}) \mathcal{S}^{-1}_j(\vec{n}). \quad (B3)$$

We can obtain the right electric field for a vertical string by parallel transporting the left electric field (32a) over string $\mathcal{U}(m,n;\hat{2})$;

$$\mathcal{E}_-(\vec{n};\hat{2}) = -\mathcal{U}^\dagger(\vec{n}-\hat{2};\hat{2}) \mathcal{E}_+(\vec{n}-\hat{2};\hat{2}) \mathcal{U}(\vec{n}-\hat{2};\hat{2}),$$
$$= E_-(\vec{n};\hat{2}) + E_+(\vec{n};\hat{1}) + E_-(\vec{n};\hat{1})$$
$$+ \sum_{j=n+1}^{N} \mathcal{S}'_j(\vec{n}) E_-(m,j;\hat{1}) \mathcal{S}'^{-1}_j(\vec{n})$$
$$- U(\vec{n};\hat{1}) \sum_{j=n+1}^{N} \mathcal{S}'_j(m+1,n) E_-(m+1,j;\hat{1})$$
$$\times \mathcal{S}'^{-1}_j(m+1,n) U^\dagger(\vec{n};\hat{1}). \quad (B4)$$

Adding (B1), (B2), (B3), and (B4) all six nonlocal terms cancel out and we see that the Gauss laws in terms of the dual potential (40) are exactly same as the original Gauss laws (8).

## APPENDIX C: THE PLAQUETTE CONSTRAINTS

In this section we show that the new plaquette constraints (41) weakly commute with the Hamiltonian and therefore remain preserved under time evolution. We define the following operator

$$\mathcal{C}_{\alpha\beta}(\vec{n}) \equiv (\mathcal{U}_{\text{p}}(\vec{n}) - \mathcal{W}(\vec{n}))_{\alpha\beta}. \quad (\text{C1})$$

In (C1) $\mathcal{U}_{\text{p}}(\vec{n}) \equiv (\mathcal{U}(\vec{n};\hat{2})\mathcal{U}(\vec{n}+\hat{2};\hat{1})\mathcal{U}^\dagger(\vec{n}+\hat{1};\hat{2})\mathcal{U}^\dagger(\vec{n};\hat{1}))$. Using (44) it is easy to prove that the Kogut-Susskind electric fields rotate $\mathcal{C}_{\alpha\beta}$ from left and right as follows:

$$[E^a_+(\vec{n}';\hat{1}), \mathcal{C}_{\alpha\beta}(\vec{n})] = \delta_{\vec{n}',\vec{n}}(\mathcal{C}(\vec{n})T^a)_{\alpha\beta}$$
$$+ \delta_{\vec{n}',\vec{n}+\hat{2}}R^{ab}(\mathcal{U}^\dagger(\vec{n};\hat{2}))(T^b\mathcal{C}(\vec{n}))_{\alpha\beta}$$
$$\approx 0, \quad (\text{C2a})$$

$$[E^a_+(\vec{n}';\hat{2}), \mathcal{C}_{\alpha\beta}(\vec{n})] = \delta_{\vec{n}',\vec{n}}(T^a\mathcal{C}(\vec{n}))_{\alpha\beta}$$
$$+ \delta_{\vec{n}',\vec{n}+\hat{2}}R^{ab}(\mathcal{U}^\dagger(\vec{n};\hat{1}))(\mathcal{C}(\vec{n})T^b)_{\alpha\beta}$$
$$\approx 0. \quad (\text{C2b})$$

In (C2a) and (C2b) we have used the plaquette, string sectors canonical commutation relations (16) and (22). They show that on the constrained surface the dual Hamiltonian commutes with the plaquette constraints:

$$[H, \mathcal{C}_{\alpha\beta}(\vec{n})] \approx 0. \quad (\text{C3})$$

We also check the commutation relations of the constraints $\mathcal{C}_{\alpha\beta}(\vec{n}) = 0$ with the Gauss law constraints (48):

$$[\mathcal{G}^a(\vec{n}'), \mathcal{C}_{\alpha\beta}(\vec{n})] = -\delta_{\vec{n}',\vec{n}}[T^a, \mathcal{C}(\vec{n})]_{\alpha\beta} \approx 0. \quad (\text{C4})$$

Therefore, the plaquette constraints (41) together with the SU(N) Gauss law constraints (48) define the physical Hilbert space where the dual loop dynamics with inverted coupling is local.